\documentclass[11pt]{article}
\usepackage{graphicx}
\usepackage{latexsym}
\usepackage[left=.75in,top=1in,right=.75in,bottom=1in]{geometry}
\usepackage{amsmath}

\newcommand\independent{\protect\mathpalette{\protect\independenT}{\perp}}
    \def\independenT#1#2{\mathrel{\setbox0\hbox{$#1#2$}%
    \copy0\kern-\wd0\mkern4mu\box0}} 

\newenvironment{packed_item}{
\begin{itemize}
 \setlength{\itemsep}{0pt}
  \setlength{\parskip}{0pt}
  \setlength{\parsep}{0pt}
}{\end{itemize}}

\newenvironment{packed_enum}{
\begin{enumerate}
 \setlength{\itemsep}{0pt}
  \setlength{\parskip}{0pt}
  \setlength{\parsep}{0pt}
}{\end{enumerate}}

\usepackage{amssymb}

\usepackage{multirow}
\usepackage[numbers, sort&compress]{natbib}
\usepackage[section]{placeins}

\usepackage[font={footnotesize,it}]{caption}

\usepackage{url}

\usepackage{pdflscape}
\usepackage{arydshln}

\usepackage[dvipsnames]{xcolor}

\usepackage[normalem]{ulem} 
\usepackage[makeroom]{cancel}

\title{Sustainable East Africa Research in Community Health (SEARCH):\\ a community cluster randomized study of HIV ``test and treat"\\
using multi-disease
approach in rural Uganda and Kenya\\ \vspace{2em}
 Statistical Analysis Plan for Phase I: \\
  Health Outcomes among Adults\\ \vspace{2em}
  }
\author{Laura  B. Balzer, PhD$^1$ \\
Diane V. Havlir, MD$^2$ \\
Joshua Schwab, MS$^3$ \\
Mark J. van der Laan, PhD$^3$ \\
Maya L. Petersen, MD, PhD$^3$ \\ \vspace{2em}
for the SEARCH Study Team \\ \vspace{2em}
}
\date{November 27, 2017\\
Version 3.0 \\ \vfill
\small{ $^1$University of Massachusetts, Amherst;
$^2$University of California, San Francisco;
$^3$University of California, Berkeley}
}

\begin{document}

\maketitle

\newpage
\tableofcontents

\newpage

\section{Overview}

The Sustainable East Africa Research in Community Health (SEARCH) Study is a two-phase pair-matched cluster randomized trial, conducted in rural Kenya and Uganda. The first phase is  designed to evaluate the impact of i) annual HIV and non-communicable disease (hypertension and diabetes) population-level testing, ii) immediate antiretroviral (ART) eligibility for all HIV-positive persons, and iii) ART, hypertension (HT) and diabetes (DM) care delivered using a streamlined patient-centered model, compared to an active control consisting of i) baseline population-level  HIV and HT/DM testing, and ii) ART and HT/DM care delivered according to national guidelines.
Inclusion and exclusion criteria as well as definition of the control and intervention arms  for Phase I are reviewed below. 
%

This document provides the analytic plan for evaluating adult HIV incidence, health, and implementation outcomes for the first phase of the SEARCH Study, including:
\begin{itemize}
\item Incident HIV infection
\item HIV testing uptake and coverage
\item ART initiation and coverage
\item Plasma HIV RNA suppression
\item Mortality 
\item Tuberculosis
\item Hypertension and diabetes control
\end{itemize}
%
Primary outcome analyses will be conducted after all SEARCH communities have completed three full years of follow-up. 

 For all comparisons between intervention and control arms, we take a two-stage approach to our analysis. In Stage I, we estimate the community-level outcomes defined below. In Stage II, we compare average the community-level outcomes between intervention and control communities, accounting for the pair-matched study design and with pre-specified adjustment for baseline factors. Primary effect estimates will be reported on a relative scale (relative risk or rate), with variance estimates based on the estimated influence curve, accounting for the pair-matched design, and inference based on the t-distribution with 15 degrees of freedom \citep{Balzer2015Adaptive, Balzer2016SATE,HayesMoulton2009}. Key analytic decisions for each outcome include the choice of candidate adjustment variables and the weights given to each community \citep{Balzer2016DataAdapt}; below, we pre-specify these decisions for each outcome considered. An additional analytic option in Stage II is to ``break the match" and treat the community as the unit of independence; we implement this as a sensitivity analysis for each outcome considered.   
 
Additional explanatory and descriptive analyses are also detailed below. We will provide descriptive statistics of the baseline characteristics of study participants, stratified by region, and (among those with baseline HIV status measured) baseline HIV status. Characteristics reported will include sex, age, marital status (single, married, widowed, divorced/separated), education (no school, primary, secondary, post-secondary), occupation, household wealth index (based on principal components analysis of baseline household socioeconomic survey, as described in protocol), self-reported alcohol and contraceptive use, self-reported prior HIV testing history, stable residence ($\leq$6 months of past year outside of the community), and mobility ($\geq$1 month of past year outside of the community). 
 
 For each study outcome compared between arms, we will report formal consort diagrams as well as participant flow diagrams enumerating inclusions and exclusions for all individuals contributing to the analysis.

The document is organized as follows.
 Section (\ref{primary.1}) provides an analytic plan for measurement and estimation of three-year  cumulative incidence of HIV for each community (Stage I). Section (\ref{primary.2}) provides an analytic plan for estimation of and inference for the effect of the randomized intervention on this outcome (Stage II). Section (\ref{power}) provides corresponding power calculations under a range of scenarios. 
Section (\ref{incidence}) provides additional analyses of the HIV incidence outcome, including evaluation of the change in annual HIV incidence  over time. Section (\ref{explanatory}) provides analyses describing contextual factors potentially contributing to observed HIV incidence patterns and investigating community-level drivers of HIV transmission. 
 Section (\ref{uptake}) describes analyses of intervention uptake, including the HIV care cascade and plasma HIV RNA suppression. 
Section (\ref{health}) describes analyses of mortality, tuberculosis, and non-communicable disease outcomes, as well as ART-associated toxicities and antiretroviral resistance.
 Full analysis plans for  (i) evaluating child and maternal HIV and health outcomes, (ii) supplemental analyses to investiage HIV transmission patterns and drivers, and (iii)  evaluating economic, educational, qualitative and costing outcomes, are provided in separate documents.

\subsection{Community selection and pair-matching}\label{match}
Fifty-four candidate communities meeting the following inclusion and exclusion criteria were initially selected in Kenya and Uganda: 
\begin{packed_item}
\item[]  \textbf{Inclusion criteria:}
\item Most recent census population between 9,000 and 11,000 individuals.
\item Served by a government health center, already providing ART or a highly functioning health center at one organizational level below those generally providing ART.
\item Community leaders' consent to ethnographic mapping.
\item Accessibility to health center via a maintained transportation route.
\item Community location with sufficient distance from other potential study communities to limit contamination of intervention or control conditions (i.e. a buffer zone).
\end{packed_item}

\begin{packed_item}
\item[] \textbf{Exclusion criteria:}
\item Presence of ongoing community-based ART intervention strategies that provide treatment outside of the current in-country treatment guidelines. 
\item An urban setting, defined as a city with a population of 100,000 or more inhabitants.
\item National government not willing or opposed to support commodities needed for Community-based Health Campaign (CHC), if provided by an outside organization.
\end{packed_item}

Data on these communities were gathered with ethnographic mapping.  Of the 54 communities, the best 16 matched pairs (5 pairs in Western Uganda, 5 pairs in Eastern Uganda, and 6 pairs in Western Kenya) were selected. Communities were matched based on region, population density, occupational mix, trading centers, and migration.

\subsection{Phase I study arms}

One community in each matched pair was randomly assigned to ``Control"  and one to ``Intervention". 
\begin{packed_item}
\item[] \textbf{Control:}
\item Baseline household enumeration
\item  At baseline (Year 0)  and Year 3, Community-based Health Campaigns (CHCs) with multi-disease prevention and treatment services, including testing and referral for  HIV, hypertension (HT), and diabetes (DM) 
\item At baseline and Year 3, home-based or in-community testing for all 
CHC non-attendees 
\item ART and HT/DM care according to national guidelines
\end{packed_item}

\begin{packed_item}
\item[]  \textbf{Intervention:}
\item Baseline household enumeration
\item Annual CHCs with multi-disease prevention and treatment services, including testing and referral for HIV, HT, and DM  
\item Annual home-based or in-community testing for all 
CHC non-attendees 
\item Streamlined HIV Care: 
\begin{packed_item}
\item Immediate ART eligibility for all HIV+
\item Supported linkage
\item Rapid ART start
\item Patient-centered care
\item Enhanced retention  
\end{packed_item}
\item Preplanned HIV cascade optimization
\item HT/DM care delivered using streamlined model
\end{packed_item}
The core intervention components and the cascade optimization strategy are detailed in a separate document.
%

\section{Stage I: Estimation of three-year cumulative HIV incidence in each community} \label{primary.1}

This section is focused on estimating the three-year cumulative HIV incidence in each community. We first define the primary cohort for measurement. Then 
we formally define the Stage I target parameter (community-specific three-year cumulative HIV incidence) and provide an overview of assumptions used to identify this parameter. Finally, we describe our estimator of the community-specific cumulative incidence, which incorporates incomplete CHC attendance, tracking coverage, and right-censoring due to death or out-migration.

\subsection{Defining the HIV Incidence Cohort}
In each community, we seek to identify a cohort of baseline HIV-negative, adult residents on whom we will measure the primary outcome of HIV seroconversion. This cohort is generated through 
\begin{packed_enum}
\item Enumeration of community residents during the baseline household census
\item  Measuring HIV status and other variable at the baseline CHC 
\item Tracking and evaluation, including home-based HIV testing, of 
residents who do not attend the baseline CHC
\end{packed_enum}
We define \emph{enumerated residents} as those  who report (or are reported by a key informant) during the census as living in a household within the community-specific geographic boundaries  or those who link to an enumerated household at the time of the baseline CHC. A resident is classified as \emph{stable} if at the time of census he or she reports (or is reported by a key informant) as living outside the parish/sublocation for $\leq$ 6 months  of the preceding 12 months.
A resident is classified as an \emph{adult} if he or she is $\geq$ 15 years of age at baseline. 

We define the \textbf{HIV Incidence Cohort} as all enumerated, stable community residents who are $\geq$ 15 years of age at baseline (adults), 
 and who have documented HIV-negative serostatus by the close of baseline tracking. 
 We exclude individuals who have moved out of the community before the close of baseline tracking, as well as  individuals who have a Ministry of Health clinic or laboratory record indicating HIV care prior to their baseline HIV rapid antibody test. 

\subsection{Individual-level data} \label{IndvData}

During the census, we measure individual-level data, including age, sex, occupation, location of residence and marital status.  For individuals attending a CHC or  non-attendees who are successfully tracked,
we measure vital status, out-migration status,  HIV status,  individual-specific covariates including changes in baseline enumeration variables 
 and health outcomes as specified in the protocol.  
 This population-based testing (via the CHC with tracking) occurs annually in the intervention communities and at years 0 and 3 in the control communities.
 

To simplify notation and presentation, we define the following individual-level data for members of the HIV Incidence Cohort: 
\begin{packed_item}
\item $W$: covariates measured during the baseline census and at the baseline CHC/tracking (e.g. age, sex, marital status, occupation, education) 
\item $C$: an indicator of death or out-migration (defined below) by year 3 
\item $\Delta$: an indicator of having HIV serostatus (and other covariates) observed at either the CHC or through tracking at year 3
\item $I$: an indicator of having a confirmed HIV-positive diagnosis at year 3 testing (CHC or post-CHC tracking).
\end{packed_item}

The observed data structure for individual $i$  is thus
\begin{equation}\label{ObsData}
O_i=\left (W_i, C_i, \Delta_i, \Delta_i I_i \right).
\end{equation}
We also observe community-level covariates $E$ (including the size of the incidence cohort in each community) and the randomly assigned community treatment arm $A$. Because $E$ and $A$ are constant across a community and thus will not impact estimation of the community-level outcome, we omit these variables from our specification of the individual-level longitudinal data structure within a community.  


\subsubsection{Out-migration}\label{sec:out-migration}

When defining out-migration, we would  ideally distinguish between migration patterns  resulting in potential exposures to HIV infection primarily within the SEARCH  community (which more accurately reflect the risk were an intervention rolled out more broadly) from migration patterns resulting in potential exposures to HIV infection primarily outside the SEARCH community (which provide less relevant information about the counterfactual exposure-level of interest). Such an ideal definition is unknown, likely to vary across communities, and likely to depend on information difficult to measure in practice. Therefore,  in the primary analysis we use a definition of out-migration that attempts to avoid censoring individuals whose primary exposure continues to occur within the community, and accept that as a result we may fail to censor people whose primary exposure is outside the community. 

An individual will be defined as out-migrated at year 3 if he or she  (or a key informant if personal report is not available) reports  either of the following.
\begin{packed_enum}
\item Spending $>$ 6 months of the preceding year outside of the community.
\item Within the preceding 3 years, spending $>$12 contiguous months outside of the community.
\end{packed_enum}

Because less mobile individuals may also represent a selected subset of the population, we will conduct a secondary analyses in which we do not censor at out-migration, as detailed below.


\subsection{Target parameter: Community-specific cumulative incidence}\label{stage1.target}

In each community-specific cohort, incident HIV cases are identified by repeat testing  at year 3 (i.e. three years after the baseline testing campaign), as described above and detailed in the protocol. 
 HIV status at year 3 will not be observed for all members of the Incidence Cohort due to death, out-migration, and 
incomplete CHC attendance with incomplete success at tracking non-attendees.
Let $I(c,\delta)$ denote the counterfactual HIV infection status at year 3 under a hypothetical intervention to set censoring $C=c$ and measurement $\Delta=\delta$. We define our Stage I target causal parameter as  the probability that a member of the HIV Incidence Cohort becomes infected with HIV during three years of follow-up
under a hypothetical intervention to prevent right-censoring and ensure knowledge of HIV status at year 3: \[
\mathbb{E}[I(0,1)]  =  \mathbb{P}[I(0,1)=1] 
\] 

 In the primary analysis, death will be treated as a right-censoring event.  In secondary analyses, we will use HIV-free survival as a composite outcome. The decision not to treat death as a competing risk in the primary analysis is based on the desire to define a community-level outcome that is not a function of underlying mortality patterns. The decision not to use HIV-free survival for our primary analysis is based on the expectation that the majority of mortality in our  HIV Incidence Cohort will not be related to HIV nor will it be strongly affected by the intervention.

In the primary analysis, out-migration will also be treated as a right-censoring event. We take this approach because subjects who migrate out of an intervention community may be exposed to a higher risk of HIV acquisition than exists within the community and thereby would dilute the effect of the intervention. Further, this dilution would be less likely to occur if a comparable strategy were rolled out region-wide, diminishing generalizability to the future context of interest. 
In secondary analyses we will (i) censor only at death and (ii) evaluate the impact of the intervention on internally-derived HIV infections, as determined through 
phylogenetic analysis  and through self-reported suspected infection source among seroconversions.

We note that defining the community-specific outcome conditional on being a member of the HIV Incidence Cohort  avoids additional assumptions on factors determining baseline testing success and corresponding complexity during estimation. However, it introduces the possibility that the HIV Incidence Cohort is not fully representative of the underlying community. Our design attempts to mitigate this risk to the extent possible by using a prioritized tracking system at baseline. After  completion of initial  tracking, any age-sex strata in which $<80\%$ of enumerated, stable, adult  residents have known serostatus are targeted for additional tracking.   To investigate the representativeness of our baseline cohort, we will report descriptive statistics comparing the age, sex and geospatial distribution of subjects seen at the baseline CHC or tracked to those enumerated in the baseline census.

\subsection{Identification and estimation}\label{Identifiability.1}

The community-specific cumulative incidence $\mathbb{E}[I(0,1)]=\mathbb{P}[I(0,1)=1]$ can be expressed as function of the observed data distribution 
if the 
randomization assumption holds \citep{Robins1986}:
\begin{align*}
&I(0,1) \independent C,\Delta \mid W 
\end{align*}
This assumption allows
 censoring (by death or out-migration) and measurement (CHC attendance and tracking success at year 3) to depend on the measured baseline covariates. 
 It fails, however, if  an individual's probability of either censoring or measurement depends on his or her interim HIV status. Reliance on this identifying  assumption would also imply the  use of a full adjustment set (i.e. all baseline covariates) during estimation of the Stage I target parameter. This would result in more complex estimators with unclear benefit to overall estimator performance (bias, mean squared error, confidence interval coverage, and Type I error control) when evaluating the impact of the intervention (Stage II target parameter). (For example, use of a full adjustment set might reduce performance if bias occurs in the same direction in intervention and control arms and/or if certain covariates strongly predict  censoring and measurement but have a minimal impact on the outcome.) 
  
  In the primary analysis, we therefore rely on the following stronger identifiability assumption:
\begin{align*}
&I(0,1) \independent C,\Delta.
\end{align*}
While HIV infection between baseline and year 3  could plausibly affect subsequent censoring or measurement (resulting in a violation of these assumptions), we minimize the extent to which this bias is likely to be differential in treatment and control arms by relying on the equivalent measurement structures  in both arms. (In other words, we only use data from baseline and year 3 for estimation of the three-year cumulative incidence in both study arms.)
Further, under a range of plausible scenarios, the direction of the bias in estimates of community-specific cumulative incidence is likely to be the same in control and intervention arms, and thus result in some degree of bias cancellation for estimates of the intervention effect.  Simulations, under a range of both plausible and extreme informative measurement and censoring processes, verify this prediction  and show good confidence interval coverage and type I error control for effect estimates based on this approach. 
 
In the primary analysis, our target statistical estimand is then
\begin{equation}\label{estimand.2}
\mathbb{E}\left[I \big|C=0, \Delta=1 \right] = \mathbb{P}\left[ I=1 \big|C=0, \Delta=1 \right].
\end{equation}
Community-specific HIV cumulative incidence will be estimated as the corresponding simple empirical proportion of Incidence Cohort members who remain alive and resident in the community at year 3 with HIV status measured at year 3 testing who are confirmed to be HIV-positive at year 3 testing. 

In sensitivity analyses, we will use baseline covariates $W$ to adjust for potentially informative censoring and missingness. When implementing these secondary analyses with a full adjustment set,  we will use targeted maximum likelihood estimation (TMLE) to estimate
\begin{equation}\label{estimand.2}
\mathbb{E}_W\mathbb{E}\left[I \big|C=0, \Delta=1,W) \right],
\end{equation}
 with Super Learning for  estimation of the outcome regression $\mathbb{E}\left[I \big|C=0, \Delta=1,W \right]$ and propensity score  $\mathbb{P}[C=0, \Delta=1 |W]$ \citep{MarkBook, SuperLearner, Petersen2014_JCI}. We will also conduct additional sensitivity analyses in which we use an analogous approach to further adjust for  known interim HIV diagnosis (acknowledging that our measurement of interim diagnosis is expected to be greater in the intervention arm). Finally, while by design we expect very similar follow-up times in intervention and control communities, in further sensitivity analyses we will estimate community-specific HIV incidence rates over the three year follow-up period (using analogous data structures in intervention versus control arms and assuming incident infections occur at the midpoint between the baseline negative HIV test and confirmed seroconversion at year 3).

Additional secondary analyses will  use the annual data available in the intervention communities  to investigate HIV incidence over time and factors contributing to ongoing transmission (as detailed further below and in the supplementary analysis plan focused on drivers of transmission). In addition, we will report estimates of the three-year cumulative HIV incidence for each community, and compare characteristics of cohort members with HIV status known versus missing at follow-up year 3. %

\section{Stage II: Estimation of the intervention effect on HIV incidence} \label{primary.2}

This section is focused on obtaining a point estimate and inference for the relative difference in 3 year HIV cumulative incidence between intervention and control arms. We first describe the community-level data and implications of  the pair-matched design. Then we specify the target parameter for Stage II as the expected HIV incidence under the intervention relative to the expected HIV incidence under the control for the study communities. Primary analysis weights each community equally; sensitivity analyses will weight individuals equally. 

Next we discuss 
two estimation strategies - unadjusted and adjusted. Our primary analysis will be adjusted.

\subsection{Community-level data and adaptive pair-matching}

Given  estimates of the community-specific cumulative HIV incidence generated in Stage I, the observed data  can be simplified to the cluster-level. Let $E$ represent the baseline community-level covariates, including measures from the ethnographic mapping (e.g. region, proximity to trucking routes, occupational mix),  the census (e.g. age distribution, sex ratio, community size), and  the baseline CHC with  tracking (e.g.  HIV prevalence). The exposure variable $A$ equals 1 if the community was randomized to the intervention arm and equals 0 if the community was randomized to the control arm. The outcome $Y$ is the estimated community-specific three-year cumulative incidence of HIV (obtained from Stage I).
Thereby, the observed data for  SEARCH community $j$ can be denoted \[
O_j = (E_j, A_j, Y_j)
\]
for $j=\{1,\ldots, 32\}$. We use $J$ to denote the total number of communities in the study ($J=32$).

As described in Section \ref{match}, fifty-four  communities were identified from rural Uganda and Kenya as potential study sites.  From these candidate communities, the  $J/2=16$  pairs (5 in Western Uganda, 5 in Eastern Uganda, and 6 in Kenya) that were best matched on baseline covariates were selected.
We consider this pair-matching scheme to be adaptive, because the partitioning of the study communities into matched pairs was a  function of the baseline covariates of all candidates. This adaptive design has important implications for estimation and inference \citep{vanderLaan2012Adaptive, Balzer2015Adaptive, Balzer2016SATE}. 
%
Given the covariates of all candidate communities, the observed data can be represented as $J/2$ conditionally independent random variables: \[
\bar{O}_k = \big(O_{k1}, O_{k2} \big) = \big( (E_{k1}, A_{k1}, Y_{k1}), (E_{k2}, A_{k2}, Y_{k2})  \big) \] 
where the index $k=\{1, \ldots, 16\}$ denotes the partitioning of the candidates  into matched pairs according to similarity on their baseline covariates. Throughout, subscripts $k1$ and $k2$ denote the first and second communities within matched pair $k$.
The treatment mechanism is known; with probability 0.5, the first unit is randomized to the intervention and the second to the control. 
Throughout we assume that the baseline covariates and the intervention assignment in one community do not affect the outcome of another study community. In other words, we assume the  study communities are causally independent. 
 Self-reported residence location of suspected infection source among seroconversions, as well as linkage of nominated social network contacts across communities, will be used to evaluate the extent of possible spill-over across communities. 

\subsection{Target parameter: Incidence ratio}

Our goal in the primary analysis is to estimate the effect of the SEARCH intervention on three-year cumulative HIV  incidence for our study communities. Let  $Y_j(a)$ denote the counterfactual cumulative HIV incidence under intervention level $A=a$ for community $j$, and let \[
\psi(a) = \frac{1}{J} \sum_{j=1}^J Y_j(a)
\]
 be the empirical mean for the study communities. Then our  target of inference is the sample cumulative incidence ratio 
 \[
\frac{\psi(1)}{\psi(0)} =  \frac{ \frac{1}{J} \sum_{j=1}^J Y_j(1) }{ \frac{1}{J} \sum_{j=1}^J Y_j(0) }  
  \]
This parameter is the average  incidence under the intervention for the 32 study communities divided by the  average   incidence under the control for the 32 study communities. 
As discussed in the following sections, estimation  and inference for the sample parameter ($\frac{1}{J} \sum_{j=1}^J Y_j(a)$) are 
identical to estimation and inference for the conditional parameter
$( 1/J \sum_j \mathbb{E}\left[ Y_j(a) \mid E_j\right])$ and analogous to the population parameter $\mathbb{E}[Y(a)]$. 
%
  The distinction lies in interpretation  and inference, with estimators of the sample parameter often being less variable (more precise) than those of the population parameter \citep{Rubin1990, Imbens2004, Imai2008, Balzer2016SATE}.

\subsection{Estimation of the intervention effect}\label{InterventionEffect}

An intuitive estimator is the  average outcome among intervention units 
divided by the average outcome among the control units: 
\[
Unadj. = 
\frac{\hat{\psi}(1)}{\hat{\psi}(0) }
= \frac{\hat{\mathbb{E}}(Y|A=1) }{ \hat{\mathbb{E}}(Y|A=0)}.
\]
When the measured covariates are predictive of the outcome, this simple estimator is often \emph{inefficient} as it fails to adjust for measured covariates  (e.g \cite{Fisher1932, Cochran1957, Cox1982, Tsiatis2008, Moore2009, Rosenblum2010}).
Irrespective of how well matching is performed, there is likely to be some residual imbalance on pre-intervention determinants of the outcome  within matched pairs.  Furthermore, there are additional baseline covariates, such as baseline HIV prevalence, that are predictive of the outcome but were unavailable during the matching process.  In general, adjusting for  baseline covariates during the analysis can reduce variance without bias, even in small trials (e.g. \cite{Moore2009, Rosenblum2010}).

Therefore, for the primary analysis we will use targeted minimum loss-based estimation (TMLE)  to provide an unbiased and more efficient estimate of the intervention  effect  \cite{MarkBook, Balzer2016SATE}. For comparison, in secondary analyses we will also implement the unadjusted estimator.
The TMLE for the sample incidence ratio  is given by the following substitution estimator: \[
TMLE =  \frac{\hat{\psi}^*(1)}{\hat{\psi}^*(0)} = \frac{ \frac{1}{J} \sum_{j=1}^J \hat{\mathbb{E}}^*(Y_j|A_j=1, E_j) }{\frac{1}{J} \sum_{j=1}^J \hat{\mathbb{E}}^*(Y_j|A_j=0, E_j)}  
\]
where  $\hat{\mathbb{E}}^*(Y|A, E)$ denotes a targeted estimate of the conditional mean function $\mathbb{E}(Y|A,E)$. In general, this targeting step is used to achieve the optimal bias-variance trade-off for the parameter of interest and to solve the efficient score equation  \citep{MarkBook, vanderLaan2006}. 
Informally, this targeting step incorporates information in the known or estimated exposure mechanism $\mathbb{P}(A|E)$. 
The algorithm is detailed in \cite{MarkBook}.

 Our \emph{a priori}-specified  library of candidate estimators of the expected outcome given intervention arm and covariates, $\mathbb{E}(Y | A, E)$,  consists of community-level logistic regressions, each with an intercept, a main term for the exposure, and either one additional covariate (baseline HIV prevalence or baseline male circumcision coverage), or no additional variable (unadjusted estimator).
 Our   \emph{a priori}-specified  library of candidate estimators of the known exposure mechanism $\mathbb{P}(A|E) = 0.5$ is defined by  logistic regression models with an intercept and a main term for one remaining covariate. For example, if baseline prevalence is selected as the adjustment variable in $\mathbb{E}(Y | A, E)$, the corresponding logistic regression is removed from the set of candidate adjustment variables for the  exposure mechanism $\mathbb{P}(A|E)$. To further reduce the library size, we  restrict the candidates  such that if the unadjusted estimator is selected for estimation of $\mathbb{E}(Y | A, E)$, we will also not adjust when estimating the known exposure mechanism.   We will use leave-one-pair-out cross-validation to select the candidate TMLE, with candidate selected to  minimize the estimated variance (described in the following section). The procedure is detailed in \citep{Balzer2016DataAdapt}.
 
%

In anticipation that certain communities may be poorly matched on baseline drivers of incidence, a challenge only partially addressed by covariate adjustment, we will also conduct two pre-specified secondary analyses: excluding the matched pair with the highest discrepancy on baseline prevalence and excluding the matched pair with the highest discrepancy on baseline male circumcision coverage. Standard power calculations suggest that the reduction in matched pair coefficient of variation $k_m$ offsets the loss in the number of independent units and thus degrees of freedom (Figure~\ref{Fig:DropPair})  \citep{HayesMoulton2009}. 
Finally, as a sensitivity analysis, we will ``break the match'', treating the community as the independent unit and including region in the pre-specified candidate adjustment variables. 

\subsection{Inference}\label{Stage2inference}

As established in \cite{Balzer2016SATE}, both the unadjusted estimator and  TMLE are asymptotically linear and normally distributed estimators of 
 treatment-specific mean: $\psi(a) = \frac{1}{J} \sum_{j=1}^J Y_j(a)$.
 %
Thus, the limit distribution of the standardized estimator is normal with mean 0 and variance given by the variance of its influence curve. 
 Under pair-matching, an asymptotically conservative approximation of the influence curve for the TMLE is given by \begin{align*}
 \hat{IC}(a; \bar{O}_k) &= \frac{1}{2} \left[  \hat{IC}(a; O_{k1}) +  \hat{IC}(a; O_{k2}) \right] \\
 \text{ with } \hat{IC}(a; O_j ) &= \frac{\mathbb{I}(A_j=a)}{ \hat{\mathbb{P}}(A_j=a |E_j)} \left( Y_j -  \hat{\mathbb{E}}^*(Y_j|A_j=a, E_j) \right)
\end{align*}
For the sample incidence ratio $\psi(1)/\psi(0)$,  we apply the Delta method to test the null hypothesis and create 95\% confidence intervals on the log-scale: \[
log[ \psi(1) / \psi(0)] = log[\psi(1)] - log[\psi(0)] 
\]
The influence curve for the $log$(TMLE) =$log[\hat{\psi}^*(1) / \hat{\psi}^*(0)]$  for matched pair $k$ is given by
\begin{align*}
\hat{IC}_{tmle}(\bar{O}_k) &= \frac{1}{\hat{\psi}^*(1)} \hat{IC}(1; \bar{O}_k) - \frac{1}{\hat{\psi}^*(0)} \hat{IC}(0; \bar{O}_k) 
\end{align*}
The unadjusted estimator is a specific case of  TMLE, where we replace the targeted $\hat{\mathbb{E}}^*(Y|a, E)$ with the empirical  $\hat{\mathbb{E}}(Y|a)$ and the estimated exposure mechanism $\hat{\mathbb{P}}(a|E)$ with empirical $\hat{\mathbb{P}}(a)=0.5$. 

Inference for the intervention effect will be based on the estimated influence curve and the  Student's $t$-distribution with 15 degrees of freedom. 
Specifically, on the log-scale we will estimate the variance by taking the sample variance of the estimated influence curve divided by $J/2$, construct Wald-Type  confidence intervals, and test the null hypothesis of no average effect. Confidence intervals and two-sided hypothesis testing will be conducted at a 5\% significance level.  Finally, the (log) point estimate and confidence intervals will be exponentiated to be on the original scale. 
While a single hypothesis test will be conducted on the relative scale, effect measures and corresponding confidence intervals will  also be reported  for the absolute  scale ($\psi(1) - \psi(0)$) to facilitate alternative uses.
Finite sample simulations suggest that under plausible scenarios the adjusted estimator provides modest to substantial efficiency gains and corresponding power improvements, while retaining good type I error control and 95\% confidence interval coverage.

%

\section{Power Calculations and Simulation Results for Primary Outcome}\label{power}
We first present standard power calculations for cluster randomized trials under a range  of plausible and conservative assumptions. 
Then in Section \ref{math_model}, we present full simulations evaluating the performance of our proposed two-stage estimator. We  report  the attained power under a range of scenarios for changes in the guidelines for ART initiation in the control arm and achieved ART coverage. We also demonstrate good 
confidence interval coverage and type I error control.

\subsection{Classical power calculations}

\begin{figure}[htbp]
\begin{center}
\includegraphics[width=0.5\textwidth]{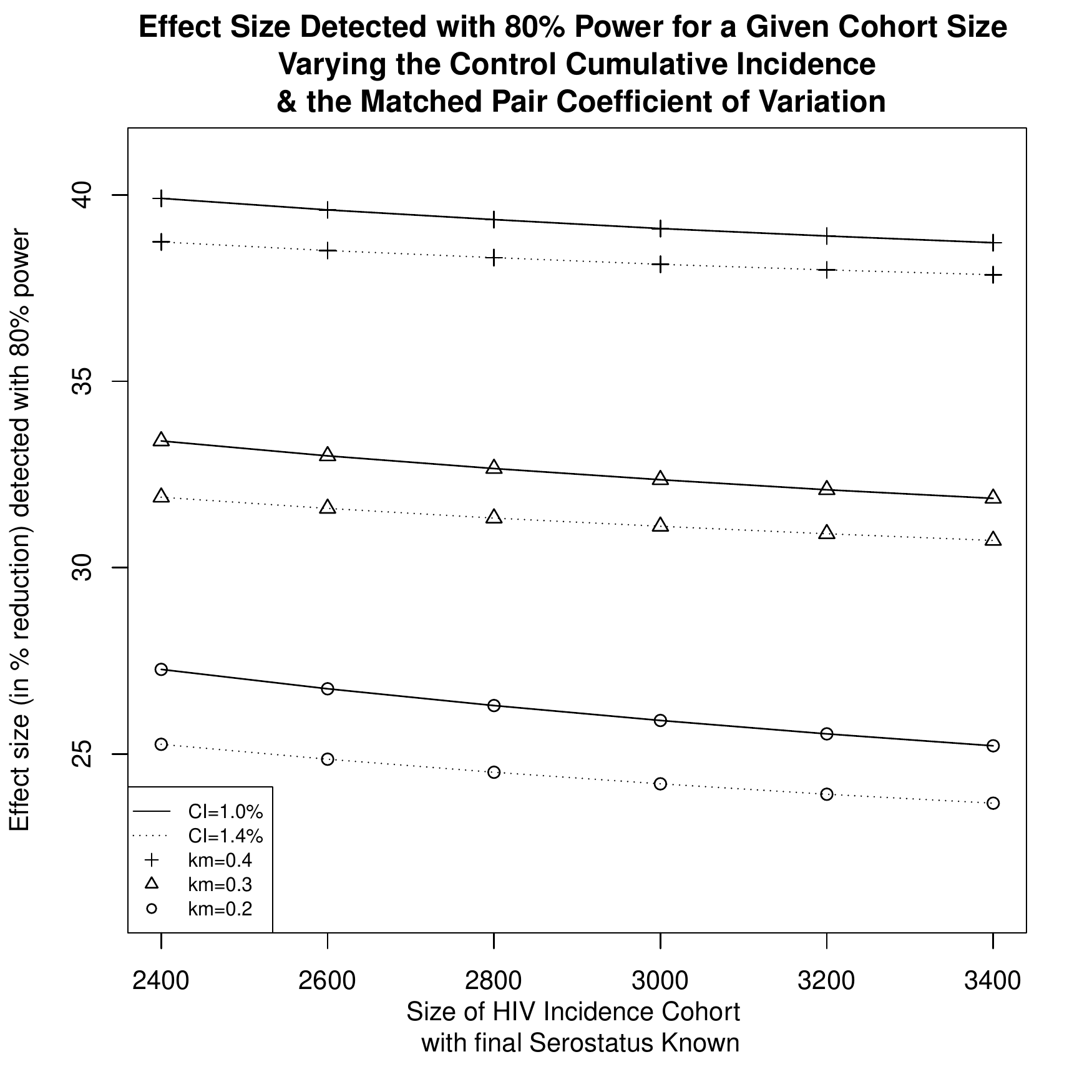}
\caption{Effect size (in percent reduction) that we are powered to detect at 80\%, while varying the control cumulative incidence (CI), the matched pair coefficient of variation $k_m$, and the number of individuals in the HIV Incidence Cohort, who have their status known at year 3. The calculations  were based on the standard sample size formulas for an unadjusted comparison of proportions in a pair-matched  cluster randomized trial with 32 communities total \citep{HayesMoulton2009}.}
\label{Fig1:HayesMoulton}
\end{center}
\end{figure}

\begin{figure}[htbp]
\begin{center}
\includegraphics[width=0.5\textwidth]{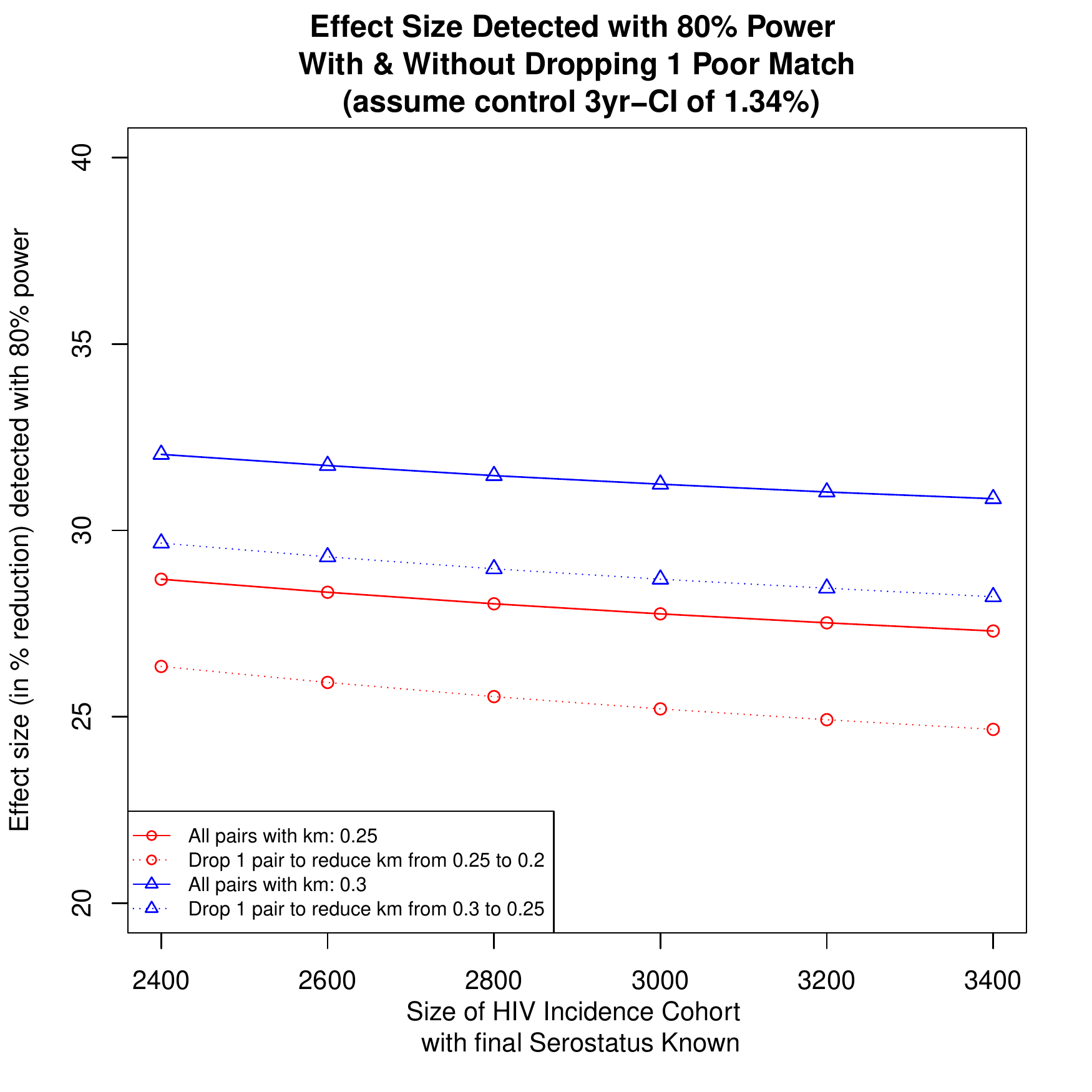}
\caption{Effect size (in percent reduction) that we are powered to detect at 80\% using all 16 matched pairs or dropping 1 poorly-matched pair resulting in a 0.05 reduction in the matched pair coefficient of variation $k_m$. The calculations  were based on the standard sample size formulas for an unadjusted comparison of proportions in a pair-matched  cluster randomized trial with two arms \citep{HayesMoulton2009}.}
\label{Fig:DropPair}
\end{center}
\end{figure}

Our initial power calculations were based on the standard sample size formulas for an unadjusted comparison of proportions in a pair-matched  cluster randomized trial with two arms \citep{HayesMoulton2009}. Using a two-sided test at a 5\% level of significance, these calculations indicated that 16 matched pairs would provide at least 80\% power to detect a 40\% reduction in the three-year HIV cumulative incidence under a conservative value for the matched pair coefficient of variation ($k_m$) and to detect smaller effect sizes under more plausible $k_m$ values. Figure~\ref{Fig1:HayesMoulton} shows a graph of the percent reduction detectable with 80\% power under a range of deviations from the following assumptions. 
 \begin{itemize}
\item     We assumed a stable adult resident size of 5,000, a baseline HIV prevalence of 10\%, measurement of HIV status at baseline among 80\% of residents, and measurement of HIV status at the final year on 75\% of 
those HIV-negative at baseline. This yields approximately 2700 residents in each community who are in the HIV Incidence Cohort  and have their serostatus known at year 3. While the exact cohort size will vary, if the actual sample size per community is at least 2700 individuals, then  these calculations can be considered conservative. We further note that moderate deviations from this number of individuals are not expected to have strong impacts on power.
\item We  assumed that the three-year cumulative HIV incidence was 1\% in  control communities. This estimate was considered conservative given the available literature, which suggested that HIV transmission rates are approximately 0.5\% to 2\% \citep{Guwatudde2009, Shafer2008, Gray2007}. For example, assuming a current incidence density of 0.5 cases per 100 person-years and allowing for a 10\% decline in transmission rate per year in the absence of the intervention (due to concurrent prevention activities and expansion of ART) would suggest a three-year cumulative incidence of approximately 1.34\%. If the three-year cumulative incidence is {\color{black}1.34\%} in control communities, then these calculations can be considered conservative. 
\item We  assumed a matched pair coefficient of variation $k_m$ of no greater than 0.4. While ideally external data would be available to the inform its selection, the generalizability of $k_m$ values across studies is limited. Specifically,  $k_m$  depends (among other things) on which covariates are used for matching, how close a match is achieved, and the strength of association between these covariates and the outcome. Furthermore, recent work has demonstrated the instability of estimates of $k_m$  based on empirical data \citep{Pagel2011intracluster}. Prior studies, performed in similar settings, have assumed a $k_m$ closer to 0.25 (e.g. Project ACCEPT [personal communication] and the Mwanza Trial  \citep{Hayes1995}). With the above assumptions, these calculations indicated that would be powered to detect a 40\% reduction with $k_m=0.4$, a 33\% reduction with $k_m=0.3$, and a {\color{black}27}\% reduction with $k_m={\color{black}0.2}$.
\end{itemize}

We also expect that these calculations are   conservative because of  the precision gained through covariate adjustment during the analysis.  Adjustment with TMLE should improve power by reducing the variability of the estimator and resulting in a less conservative variance estimator \citep{Balzer2016SATE}. 
On the other hand, these calculations may be anti-conservative if there is substantial heterogeneity in HIV incidence within study regions and we match poorly on those sources of variability within region.

\FloatBarrier
\subsection{Simulations and mathematical modeling}\label{math_model}

We used simulations to examine the performance of our proposed two-stage effect estimator. We first used a mathematical model to generate plausible country-specific incidence curves under a range of assumptions regarding scale-up of ART. These incidence curves were the basis for full hierarchical simulations that incorporated a number of the challenges faced by our primary outcome analysis, including: 
(i) differential {\color{black}HIV testing} processes in the intervention and control arms,  (ii) possibly differential informative right-censoring (due to death and out-migration) by intervention arm and  individual HIV status, (iii) possibly differential informative measurement (through CHC attendance and tracking success) by intervention arm and individual HIV status, (iv)  the inability to match on all measured baseline covariates predictive of the outcome, (v) few conditionally independent units, and (vi) rare outcomes.

\subsubsection{Simulation setup}

The following describes the data generating experiment for each of the 32 communities in the simulated study. 
To reflect the underlying processes, including differential   HIV testing between study arms, we simulated the complete data at $t=\{0,1,2,3\}$ for all individuals in each community. 
As previously discussed, our estimators only use data measured at $t=\{0,3\}$ in both arms.
 We first describe the generation of the community-level data and then  the individual-level data. Throughout,  we use $i$ to denote individuals in community $j$ at time $t$.

For community $j$, nine baseline community-level covariates were generated by drawing from a multivariate normal. 
The correlation between  the first three covariates $\{E1_j, E2_j, E3_j\}$ and between the second three covariates $\{E4_j, E5_j, E6_j\}$ was approximately 0.25, while the correlation between the last three $\{E7_j, E8_j, E9_j\}$ was 0.
Region $R_j$ was set to reflect the study design with 10  communities from Eastern Uganda, 10 communities from South Western Uganda,  and 12 communities from Kenya. Pre-intervention HIV prevalence $Z_j$ was generated  to reflect  baseline study data and as a function of region $R_j$,  covariates $\{E1_j, E4_j,  E7_j\}$, and random noise $U_{Z_j}$. Baseline coverage of male circumcision $Z2_j$ was also generated to reflect baseline study data. 

 The community-specific hazard of HIV infection under study arm $a$ at time $t$, denoted $h_{jt}(a)$, was  generated as a function of the projected incidence rate\footnote{The incidence rate of HIV under exposure-level $A=a$ at time $t$ was informed by Goals module from the Spectrum System of Futures Institute, as detailed in Section~\ref{MathModel}.},  community covariates $\{E2_j, E5_j, E8_j\}$, prevalence $Z_j$, circumcision coverage $Z2_j$, and  random noise that was correlated within a community over time. 
The number of stable, adult residents  was drawn from a uniform with minimum 4,000 and maximum 6,000 for Ugandan communities and with minimum 3,500 and maximum 5,480 for Kenyan communities.
The baseline coverage of HIV testing (via the baseline CHC and tracking) was drawn from a uniform with minimum of 80\% and maximum of 90\%.

For individual $i$ in community $j$ at time $t=0$, baseline HIV status $Y_{ij0}$ was generated as a function  the baseline community-level prevalence $Z_j$ and  random noise that was correlated within an individual over time  $U_{Y_{ijt}}$.
Baseline measurement  (CHC attendance or post-CHC tracking) $\Delta_{ij0}$ was  generated as a function of the community-specific coverage probability at baseline and random noise  
that was correlated within an individual over time $U_{\Delta_{ijt}}$. The resulting HIV Incidence Cohort was then defined as all community members who were HIV-negative and observed at baseline: $Y_{ij0}=0$ \& $\Delta_{ij0}=1$. (All community members were assumed to be living, stable residents at baseline: $C_{ij0}=0$ for all $i$ and $j$.)

For the remaining years of the trial $t>0$,  HIV status $Y_{ijt}$  was generated as  a function the community-specific hazard  $h_{jt}(a)$, individual-level covariates of age, sex, and circumcision (among males), and random noise $U_{Y_{ijt}}$. 
(The individual-level covariates were also generated to reflect baseline data.) Censoring, representing both death and out-migration, was generated as a function of the study arm $A$, underlying HIV status $Y_{ijt}$, and random noise that was correlated within an individual over time $U_{C_{ijt}}$. 
For simplicity, we assumed that past measurement  
did not affect censoring at $t$.
We explored a variety of censoring mechanisms, ranging from non-differential to  quite differential by study arm and underlying HIV status. We also explored a ``mixture" scenario, where each community was randomly and independently assigned a censoring scenario with equal probability.  The mixture scenario reflects that censoring might be operating in different ways in different communities. 

We also explored two measurement (CHC attendance and tracking) mechanisms. In the first, the observation status after baseline $\Delta_{ijt}$ (for $t>0$) was generated as function of the study arm $A$, underlying HIV status $Y_{ijt}$, censoring $C_{ijt}$, and random noise $U_{\Delta_{ijt}}$. In the second,  the observation status $\Delta_{ijt}$ was generated as a function of the study arm $A$, known HIV+ status, censoring $C_{ijt}$, and random noise $U_{\Delta_{ijt}}$. Here, HIV+ status was ``known" if an individual tested positive at a prior CHC or subsequent tracking. 
 For each type of measurement mechanism (i.e. dependent on underlying HIV status or ``known" HIV status), we explored a variety of scenarios, ranging from non-informative to quite informative  by HIV status and treatment arm. As before, we generated a ``mixture" scenario, where each community was randomly and independently assigned a  measurement scenario with equal probability.  
 By definition, the observation probability was 0 for control community members at $t=\{1,2\}$.

Given simulated data under both study arms, we calculated as the true value of our  target parameter,  the sample incidence ratio:
 \begin{align*}
\frac{\psi(1)}{\psi(0)} &=  \frac{ \frac{1}{J} \sum_{j=1}^J  Y_j(1)}{\frac{1}{J} \sum_{j=1}^J  Y_j(0)} \end{align*}
where $Y_j(a)$ denotes the three-year cumulative HIV incidence for community $j$ under the exposure-level $(A=a)$ and under a hypothetical intervention to prevent censoring and ensure final knowledge of HIV status among all members of the community-specific HIV Incidence Cohort.
 
\subsubsection{Adaptive pair-matching, intervention randomization, and estimation}

Using the non-bipartite matching algorithm \texttt{npbMatch}  \citep{nbpMatching}, we  pair-matched communities within region $R$ on predictors of baseline prevalence $\{E4, E7\}$. 
The  intervention $A$ was randomized within the matched pairs. 
For Stage I estimation of the community-specific cumulative incidence of HIV, we implemented the unadjusted estimator based on a simple empirical mean. 
For Stage II estimation of the intervention effect, we implemented both the unadjusted estimator as well as the  pre-specified data-adaptive procedure, described in Section~\ref{InterventionEffect}.   
  Inference was based on the estimated influence curve. For confidence interval construction and two-sided hypothesis testing, we used Student's $t$-distribution with 15 degrees of freedom and a 5\% significance level.
These estimation procedures were previously detailed in Sections \ref{primary.1} and \ref{primary.2}. 
	

\subsubsection{Mathematical modeling} \label{MathModel}

The Goals module from the Spectrum System of Futures Institute was used to provide  country-specific projections of the prevalence and incidence of HIV under the SEARCH intervention and under the control (\url{http://www.futuresinstitute.org/spectrum.aspx}). The software was originally developed by the Futures Group, in collaboration with Family Health International, and is supported by UNAIDS and the Gates Foundation, among others \citep{SpectrumManual, GOALSmanual, AIMmanual2014}.   
	
\begin{figure}[htbp]
\begin{center}
\includegraphics[width=0.45\textwidth]{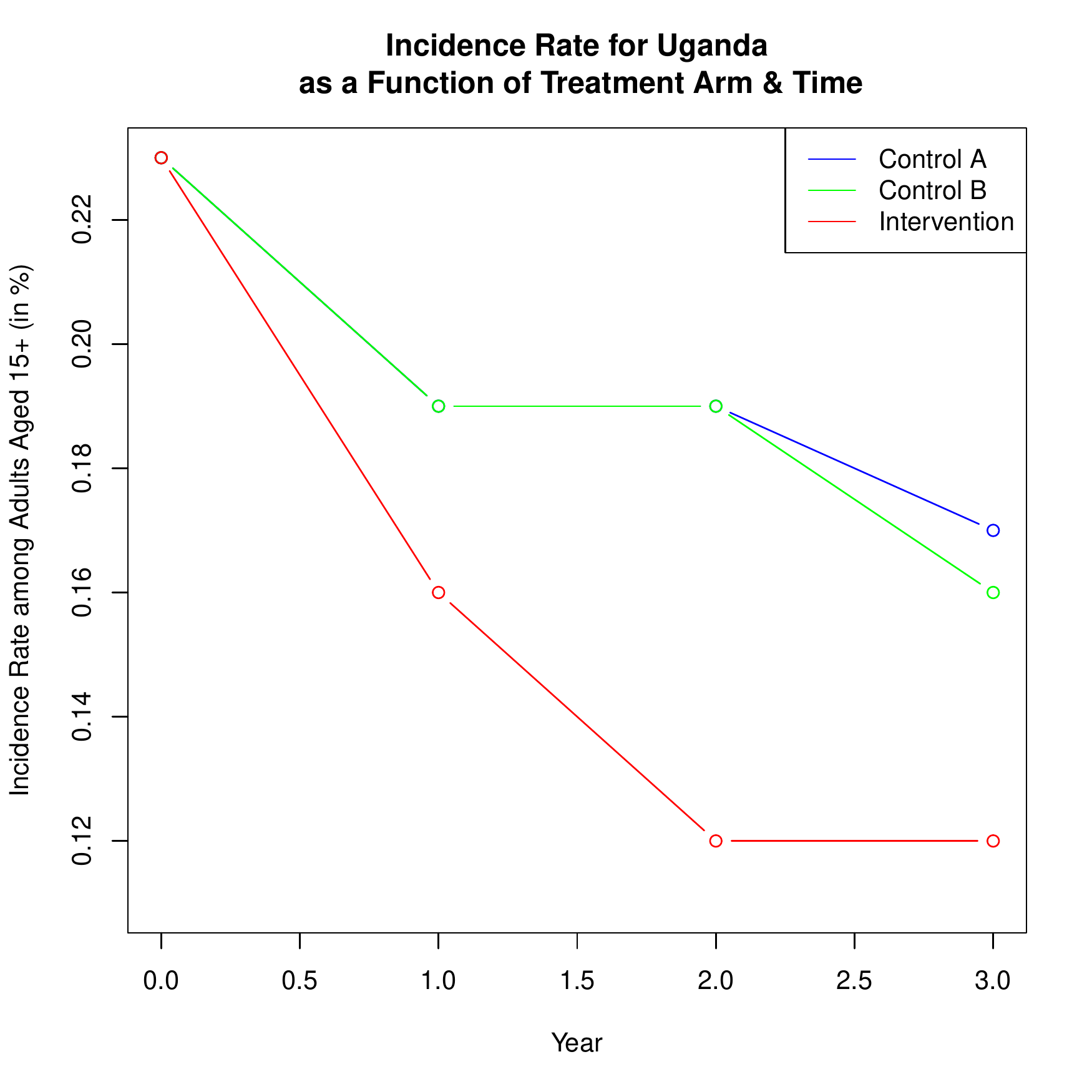}
\includegraphics[width=0.45\textwidth]{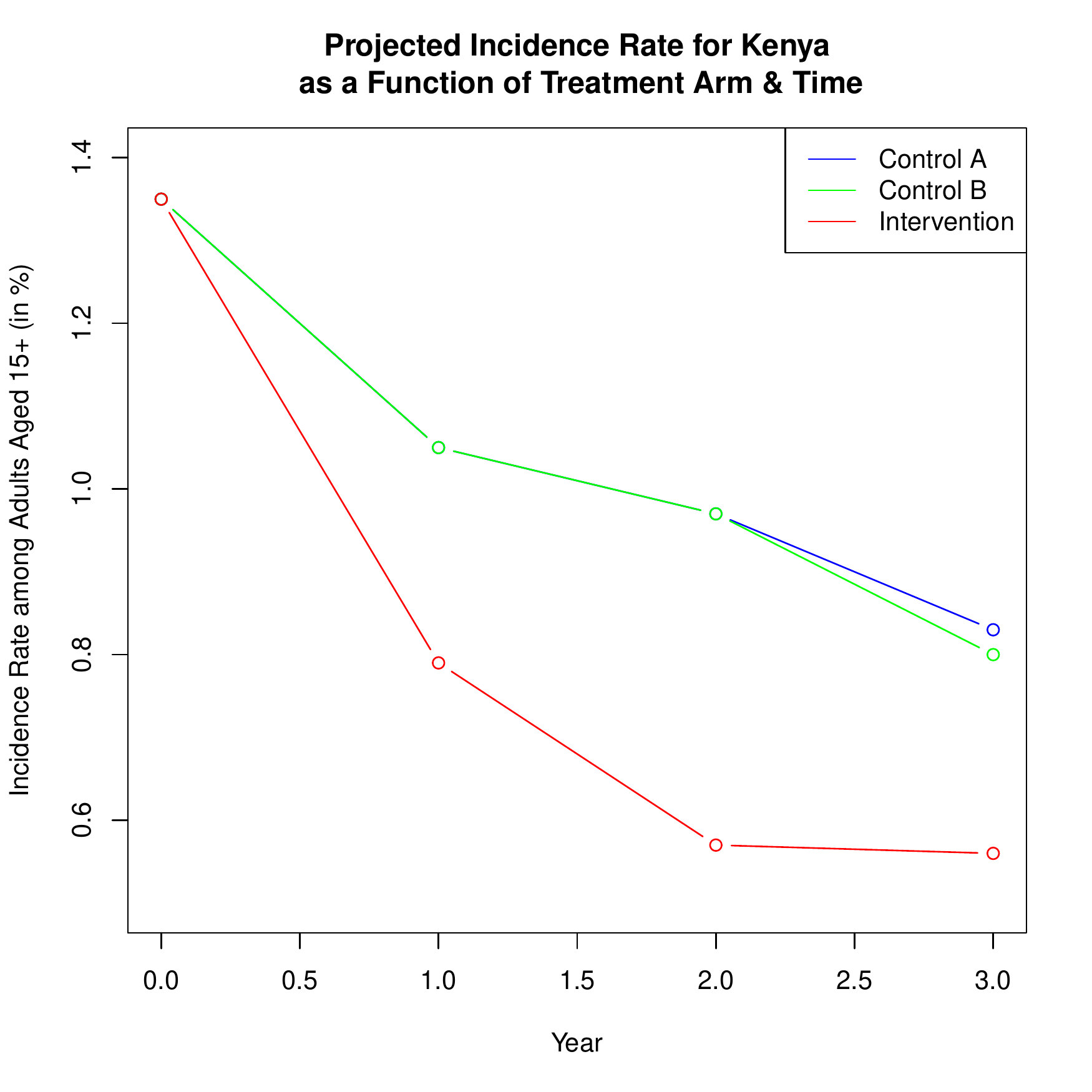}
\end{center}
\caption{Projected incidence of HIV for  Uganda and for  Kenya  in percent (\%) as informed by the Goals module in the Spectrum System of Futures Institute \citep{SpectrumManual, GOALSmanual, AIMmanual2014}. The projected incidence rates (in percent) under the control scenario A (less conservative), control scenario B  (more conservative) and the intervention arm are given by blue, green and red lines, respectively. }
\label{Fig:MathModel}
\end{figure}

The mathematical model was parameterized with published  data from national and regional surveys in Uganda and Kenya on HIV prevalence and coverage of male circumcision (Nyanza region of Kenya) \citep{UHSBS2004, Shafer2008, UAIS2011, UNAIDS2013, KAIS2014, Kimanga2014}.  
At baseline, we assumed 75\% of eligible populations  in Uganda and 66\% in Kenya were on ART and virally suppressed. 
The model was also parameterized to reflect post-baseline changes in ART eligibility 
 as well as scale-up of ART coverage. Specifically, the inputs for the control arm reflected in-country implementation of the  guidelines to change CD4-based eligibility from $\leq 350$ cells$/\mu L$ to $\leq 500$ cells$/\mu L$  starting in 2014; universal eligibility for key populations, including pregnant women, tuberculosis/HIV co-infected and discordant couples, starting in 2015{\color{black}; and universal eligibility for all HIV+ starting in  2016.}
We generated incidence curves in the control arm under these guidelines and a range ART coverage trajectories (control scenarios A-B), in which 62-67\% of eligible populations under expanded guidelines were on ART and virally suppressed by year 3 of the study. 
  These were then contrasted with the projected  incidence curves under the SEARCH intervention, assuming 73\% of all HIV+ were on ART and suppressed by 18 months after baseline (Figure~\ref{Fig:MathModel}).

\subsubsection{Results}

The true value of the sample effects depends on the $n = 32$ communities in the study. Over the 500 simulated data sets, Table~\ref{Table:Truth} shows the average value of the sample incidence ratio: $\psi(1)/\psi(0)$. The variance of the effect and the average matched pair of coefficient of variation $k_m$ are also given.  The scenarios explored are described in Figure~\ref{Fig:MathModel}.

\begin{table}[ht]
\centering
\begin{tabular}{l | rrr}
  \hline
 &  $\psi(1)/\psi(0)$ & $Var[ \psi(1)/\psi(0)]$ & $k_m$ \\ 
  \hline
Scenario A & 0.695 & 7.04E-5 & 0.372 \\ 
  Scenario B & 0.704 & 7.55E-5 & 0.366 \\ 
\hline
\end{tabular}
\caption{The average values and variance of the causal parameter across 500 repetitions of the data generating experiment.   $\psi(1)$ refers to the average risk (three-year cumulative incidence of HIV) under the intervention, and $\psi(0)$ refers to the average risk under the control. Recall this parameter changes with each sample, and $Var[\psi(1)/\psi(0)]$ gives the variability of the effect across the 500 runs. Finally $k_m$ is the average value of the estimated matched pair coefficient of variation.} 
\label{Table:Truth}
\end{table}

Table~\ref{Tab:Phase1}   illustrates the performance of the estimators over 500 simulated data sets.   Specifically, we compare the unadjusted estimator and the TMLE.  Both estimators were unbiased.   As expected, there was an efficiency gain with adjustment through TMLE.
The attained power ranged from 81 to 83\% with the unadjusted and from 88 to 91\% with the TMLE.
Throughout, there was nominal to conservative confidence interval coverage and type I error control (Table~\ref{Tab:Phase1}). 
Simulation results under alternative matching schemes, differential censoring, and informative missingness are available elsewhere.

\begin{table}[ht]
\centering
\begin{tabular}{l | rrrrr|rrrrr}
  \hline
 & \multicolumn{5}{c}{\emph{Unadjusted Estimator}} & \multicolumn{5}{c}{  \emph{TMLE}} \\ \hline
 & bias & std. err. & tstat & cover & power & bias &std. err.  & tstat & cover & power  \\ 
 Scenario A & 1.46E-3 & 1.34E-2 & -3.31 & 0.98 & 0.83 & 2.45E-4 & 9.85E-3 & -3.91 & 0.95 & 0.91 \\ 
  Scenario B & 2.10E-3 & 1.34E-2 & -3.18 & 0.97 & 0.81 & 1.28E-3 & 1.01E-2 & -3.73 & 0.95 & 0.88 \\ 
  \hline
   & bias & std. err. & tstat & cover & $\alpha$ & bias &std. err.  & tstat & cover & $\alpha$ \\ 
        Null & 5.60E-3 & 1.31E-2 & -0.00 & 0.97 & 0.03 & 5.06E-3 & 9.30E-3 & -0.00 & 0.95 & 0.05 \\ 
   \hline
\end{tabular}
\caption{The bias (average deviation between the point estimate and sample-specific true value), average standard error (estimated with the influence curve), average value of the test statistic (point estimate divided by standard error estimate), confidence interval coverage (proportion of intervals containing the true parameter value),  attained power (proportion of studies correctly rejecting the false null hypothesis), and type I error rate $\alpha$  (proportion of studies falsely rejecting the true null hypothesis) of the unadjusted estimator and the TMLE over 500 simulated trials. The null scenario was simulated by generating incidence as if the intervention had 0 impact (i.e. the hazard of HIV infection under the intervention equaled that of the control).}
\label{Tab:Phase1}
\end{table}



%

\section{Additional analyses of incident HIV infection}\label{incidence}

\subsection{Subgroup analyses for the primary outcome}
 
For the following baseline subgroups, we will report estimates of the  three-year cumulative HIV incidence by treatment arm (conducted as for Stage I of the primary analysis) and estimate the effect of the randomized intervention on this outcome, including  a formal hypothesis test of no intervention effect (conducted as for Stage II of the primary analysis): sex, age (15-24 years and $>$ 24 years), marital status (ever vs. never married), non-mobile populations ($<1$month of past year spent away from the community), and uncircumcised men. We will further compare intervention versus control incidence for each region separately; Stage II analyses for region-specific comparisons will be unadjusted due to the small number of communities.  

In addition, for the following baseline strata, we will report estimates of the three-year cumulative HIV incidence by treatment arm:
 adolescents  (15-24 years; overall and by sex), adults aged 15-49 years, adults aged 15-59 years,
mobile populations ($\geq$1month of past year away from the community), 
non-stable residents ($>$6month of past year away from the community),
circumcised men, students, and fishermen. 
%
%

\subsection{Change in HIV incidence over time}
To understand the changes in HIV incidence over time, we will estimate the annual incidence of HIV in the intervention arm, making use of community-based testing results from annual CHC and tracking to construct three annual incidence cohorts. Specifically, for testing years $t \in\{0,1,2\}$ (where $t=0$ denotes study baseline), among an open cohort of adult (aged $\geq15$ years at year $t$) residents (including inmigrants identified through the follow-up year 3 re-census) who test HIV-negative at year $t$, we will estimate  HIV incidence during the subsequent year of study follow-up (infection by testing round $t+1$). Estimates will be reported overall, and further stratified by region, gender, age ($\leq 24$, $>24$), and circumcision status, as well as reported for each community. 

Primary analyses will report estimates of annual HIV incidence rate among individuals with measured HIV status at year $t+1$, assuming incident infections occur at the mid-point between negative and positive tests, excluding individuals who have out-migrated during the year. Sensitivity analyses will (i) not censor at outmigration; (ii) restrict to baseline stable residents (acknowledging the potential depletion of high risk individuals, even after adjustment for measured differences); (iii) calculate annual risks vs. rates 
using a 95-day annual testing window;
and (iv) adjust for potentially informative incomplete ascertainment of HIV status (including due to censoring by death or outmigration in the interim year). 



The primary analysis to evaluate the change in annual HIV incidence rate over time will use Poisson regression with generalized estimating equations, and with standard errors estimated using an robust sandwich estimator based on an 
an exchangeable working covariance matrix. 
We will estimate change over time both with and without adjustment for changes in the characteristics of the measured incidence cohort.  

We will further estimate the change in annual risk of HIV acquisition using a pooled individual interval-level TMLE, with the exposure of interest defined as the time interval, and influence curve-based estimation of the standard error, respecting the community as the independent unit (i.e. allowing for dependence of observations within a community conditional on the covariates included in the adjustment set), and using the t-distribution as the basis for statistical inference. 

\subsection{Individual-level predictors of HIV seroconversion}

We will provide descriptive statistics of members of the HIV Incidence Cohort who seroconverted by year 3, including the distributions of age, sex, and other baseline characteristics, stratified by arm and by region. For each of the following baseline predictor variables, we will report unadjusted associations and adjusted variable importance measures on the relative scale (statistical analogs of the causal risk ratio), treating each baseline predictor in turn as the intervention variable, and the remainder (together with region) as the adjustment set. 
Baseline predictors to be considered will include sex, age, marital status, education, occupation, household wealth index, mobility, circumcision (among men), self-reported alcohol use, self-reported contraceptive use, relationship to head of household, polygamy, self-reported prior HIV testing, and baseline testing location (CHC vs. tracking).

Variable importance measures will be estimated with pooled individual-level TMLE with the community treated as the independent unit for influence curve-based variance estimation and with inference based on the Student's t-distribution. Secondary analyses will treat individuals as the independent unit and include community as a fixed effect.
Variable importance measures will be calculated with and without adjustment for potentially informative censoring and measured HIV status, and will be reported overall and stratified by arm, region and sex. 

\subsection{Potential sources of HIV seroconversions and internally-derived HIV infections}

\subsubsection{Discordant spouses}

We will provide descriptive statistics of known discordant couples at baseline. For the overall pooled HIV Incidence Cohort (pooled over all communities and arms) and each arm-specific HIV Incidence Cohort (pooled over all communities within each arm), we will report the number and proportion of seroconversions that occurred among baseline discordant couples. 


\subsubsection{Seroconversion interviews}

Based on qualitative interviews of participants who seroconverted, we will create descriptive tables of the self-reported suspected source of HIV infection. Using self-reported residence  of suspected infection source, we will classify seroconversions as  internal or external to the seroconverter's community of residence (or ``unable to classify"). 
We will estimate an alternative community-level HIV incidence outcome, defined as the probability of becoming infected over the 3 years of the study by a suspected source resident in the same community (``internally derived" by self-report).  We will conduct analyses analogous to those performed for the primary outcome to compare the probability of becoming infected by an internally derived virus (with and without including seroconversions with unknown source) across treatment arms and among subgroups. 

\subsubsection{Phylogenetic analyses}
Viral consensus sequences will be used to estimate phylogenetic relationships and genetic distances between HIV viruses sampled  during the study. These data, together with additional reference sequences, will be used to classify incident HIV infections among community cohort members as linked or not linked to previously documented infections among community members \cite{Campbell2011}. 

An {internally-derived} infection will be defined as a seroconversion 
  classified, based on sequence analysis, as linked to a  previously measured virus from a member of the same community.  An {externally-derived} infection will be defined as a seroconversion 
 classified as unlinked to a previously measured virus from a member of the same community.  We will estimate an alternative community-level HIV incidence outcome, defined as the probability of an HIV Incidence Cohort member becoming infected over the 3 years of the study by an internally-derived virus.  We will conduct analyses analogous to those performed for the primary outcome to compare the probability of becoming infected by an internally derived virus across treatment arms and among subgroups. We will also provide descriptive statistics and evaluate predictors of internally-derived infection. 
  
In addition, characteristics of transmission clusters detected using phylogenetic data will be reported, including age, sex, occupation, mobility, discordant spouses, shared household membership, and geospatial proximity. 
Information from seroconversion interviews will further be used to identify possible transmission links and shared risks between cluster members.

\section{Community-level descriptive and explanatory analyses} \label{explanatory}
At baseline and year 3, stratified by intervention arm, we will estimate the following potentially important drivers of HIV incidence: 
HIV prevalence, 
male circumcision coverage (traditional, medical, and overall),
HIV RNA suppression, and migration status. 
Sexual behavior and mixing patterns will be further investigated in a complimentary nested study that will  provide more detailed measurement and analysis of mobility.

\subsection{Community-level drivers of HIV incidence }

\subsubsection{HIV prevalence}
We will report HIV prevalence at baseline and year 3 among all adult  residents.
We will estimate prevalence  by community, region, treatment arm, age-sex strata (here and throughout this section, using as age categories 15-24, 25-34, 35-44, 45-54, and $>$55 years), and baseline mobility (here and throughout this section, using as primary mobility categories those reporting $<$1mo away vs. $\geq$1mo away from the community in past year).
 We will compute both unadjusted prevalence estimates based on the empirical proportion of HIV-positive individuals among those tested, and adjusted estimates, accounting for incomplete coverage of HIV testing. For the latter, we will use TMLE to adjust for ways in which individuals tested for HIV are different from those not tested \citep{PetersenCascade2017, Balzer2017CascadeMethods}. 
 Among individuals who test HIV-positive at baseline, we will report the proportion in the following CD4 strata: $<$50 cells$/\mu l $, 50-200 cells$/\mu l$, 201-349 cells$/ \mu l$, 350-500 cells$/\mu l$, and $>$500 cells$/\mu l$.

\subsubsection{Male circumcision coverage}
At baseline and year 3, we will report the proportion of adult male residents who are circumcised (overall and  by medical vs. traditional means). We will also report coverage by community, region, treatment arm, age strata, and baseline mobility. Estimates will be based on  the unadjusted empirical proportion of circumcised adult males among those seen at the CHC/tracking and adjusted estimates, accounting for incomplete measurement with TMLE to adjust for ways in which individuals seen are different from those not \citep{PetersenCascade2017, Balzer2017CascadeMethods}.

\subsubsection{Plasma HIV RNA $\geq$500 copies/ml}
Among individuals known to be HIV-positive at baseline, we will report the proportion in the following baseline plasma HIV RNA strata: $<$500 cps/ml, 500-999 cps/ml, 1,000-9,999 cps/ml, 10,000-99,999 cps/ml, $\geq$100,000 cps/ml). 

We will  estimate the  proportion of the total adult population, and of the HIV-positive adult population, with  plasma HIV RNA $\geq$500 copies/ml at baseline and at year 3, adjusting for incomplete measures of HIV status and HIV RNA levels among HIV-positive individuals, as detailed in  Section~\ref{uptake} (``Population-level HIV RNA metrics"); sensitivity analyses will consider a threshold of 1000 copies/ml. Estimates will be reported by  community, region, treatment arm, age-sex strata, and baseline mobility. We will further use adjusted variable importance measures, estimated using a  pooled individual-level TMLE with community included as a fixed effect, to evaluate predictors of having  plasma HIV RNA level $\geq$500 copies/ml at year 3.

In addition, to investigate relationships between community-level HIV viral replication and HIV incidence, we will estimate
the proportion of cumulative person time (of adult person time in the community contributed by all residents, not only those who are HIV-positive) with unsuppressed HIV viral replication (plasma HIV RNA $>$500 copies/ml) within each community. Estimation of such a metric is complicated by different measurement structures in the intervention and the control arms. Specifically, 
 HIV status, plasma HIV RNA levels, and in-migration to the study communities are measured at only two time points in the control communities. In contrast,  in the intervention communities, HIV status is  measured annually at CHC/tracking; HIV RNA is measured both at annual CHC/tracking and during interim clinic visits, and in-migrants to the community are (partially) ascertained annually. Further, a linear extrapolation between community-specific proportions unsuppressed at baseline and year 3 will fail to detect any change in the shape of the suppression curve over time, as might be expected to result from the intervention (for example, due to expanded ART eligibility at baseline and facilitated linkage with streamlined ART delivery post-baseline in the intervention communities). 
 
 We will therefore estimate the total unsuppressed person-time for each community with an algorithm  that relies on baseline and year 3 measures of HIV status, HIV RNA level, and migration status only (to ensure comparability between arms), but also incorporates interim data on ART initiation date (ascertained in both arms).
 
For these analyses, adult person-time in a community will begin at the first of the start of the community-specific baseline CHC (for baseline residents aged $\geq15$ years at baseline) or first date at which an individual is a resident (including in-migration) and is aged 15 years old or more. 
 Adult person-time in the community will end at the first of (i) date of death, (ii) date of out-migration (if any), and (iii) end of year 3 tracking. Within this person-time, we will estimate the total person-time with unsuppressed HIV RNA levels in the population using the following algorithm.

\begin{enumerate}
\item Estimate total unsuppressed time between baseline and year 3 contributed by adult residents diagnosed with HIV at or before study baseline.
	\begin{packed_item}
	\item Assume baseline HIV-positive residents who are suppressed at baseline and suppressed at year 3 are never unsuppressed.
	\item Assume baseline HIV-positive residents who are unsuppressed at baseline and suppressed at year 3 are unsuppressed until minimum of 6 months after ART initiation date or year 3.
	\item Assume baseline HIV-positive residents who are unsuppressed at baseline and unsuppressed at year 3 are always unsuppressed.
	\item Among individuals in each of the categories above who are classified as outmigrants at year 3, censor person-time at date of out-migration.
\end{packed_item}
\item Estimate total unsuppressed time between baseline and year 3 contributed by incident HIV infections (tested HIV-negative at baseline and HIV-positive at year 3) among baseline residents
\begin{packed_item}
\item Among incident infections that are unsuppressed at year 3, assume that infection occurred midway between baseline and year 3, and the individual was never suppressed.
\item Among incident infections that are suppressed at year 3, assume that the infection occurred midway between baseline and ART initiation date, and that the individual was unsuppressed from time of infection until the minimum of 6 months after ART initiation date or year 3.
\item Among individuals in each of the categories above who are classified as outmigrants at year 3,  censor person-time at date of out-migration.
\end{packed_item}
\item Estimate total unsuppressed time between baseline and year 3 contributed by HIV-infected in-migrants (tested HIV-positive at year 3 and are not baseline enumerated residents).
\begin{packed_item}
\item Among HIV-positive in-migrants who are unsuppressed at year 3, assume that the individual was HIV-positive at date of in-migration, and that the individual was never suppressed.
\item Among HIV-positive in-migrants who are suppressed at year 3, assume that  the individual was HIV-positive at date of in-migration, and that the individual was unsuppressed from time of in-migration until the minimum of 6 months after ART initiation date or year 3. 
\end{packed_item}
\item Estimate total unsuppressed time between baseline and year 3 contributed by baseline enumerated residents who are missing baseline HIV status and who are HIV+ at year 3.
\begin{packed_item}
\item Among individuals who are missing baseline status and tested HIV-positive at year 3 (this includes both incident infections and baseline prevalent HIV-positive), assume that individual was baseline prevalent HIV-positive. If suppressed at year 3, assume always suppressed. If unsuppressed at year 3, assume never suppressed. If classified as out-migrants at year 3,  censor person-time at date of out-migration. 
\end{packed_item}
\end{enumerate}

 For analyses including the intervention arm only, we will construct analogous estimates of total non-suppressed time, incorporating all available interim data on migration, HIV status, and HIV RNA levels.
 
\subsubsection{Migration \& Mobility}
Mobility may impact health outcomes and HIV transmission risk in a number of ways. First, mobile HIV-positive individuals may be less likely to be diagnosed, treated, and virally suppressed. Thereby, mobile HIV-positive individuals may be at risk of poor health outcomes and  of transmitting HIV. 
Second, individuals who migrate into the community during the study will not have benefited from the intervention prior to moving into the community. Such individuals further are not systematically ascertained during interim years, and as a result will not be tracked if they do not attend an annual campaign and may fail to fully benefit from the  intervention.  Third, individuals who migrate out of the community, as well as those who remain official residents but spend substantial time in other locations, may have greater challenges accessing testing, treatment and care services, while nonetheless continuing sexual contact with community residents.  
Finally, mobility may also be associated with additional factors that place individuals at risk of HIV transmission and acquisition (such as occupations that involve transactional sex). Mobile individuals may also have more sexual contacts with residents of communities not served by the SEARCH intervention and thereby sexual contacts who are less likely to be virally suppressed if HIV-positive. 

We will, therefore, conduct analyses to  quantify migration, to investigate HIV care cascade outcomes among mobile individuals, and investigate the role mobile individuals play in ongoing transmission.
First, we will report descriptive statistics on the following  metrics, stratified by treatment arm, by community,  and within treatment arm by  region, sex,  and age.
\begin{itemize}
\item Baseline mobility: Among baseline adult residents, we will provide descriptive statistics on months spent outside community in the past year, an indicator of moving main residence within the past year, and  nights spent at the main residence in the past month. 

\item Follow-up year 3 mobility: Among adult residents at year 3, we will provide descriptive statistics on time at current residency (less than 1 year, 1-2 years, 2+ years), months spent outside the community in the past year, an indicator of moving main residence within the past year,  nights spent at the main residence in the past month, indicator of spending more than 6 contiguous months away from main residence in the last year, an indicator of spending more than 12 contiguous months away outside of the community in the last 3 years, and an indicator of living in the community for the past 5 years. In this population, we will further report the number and proportion who are classified as in-migrants, defined as an individual resident in the community at year three, but not at baseline enumeration.
 We will  provide basic descriptive statistics of the in-migrants' characteristics (e.g. sex, age, occupation, relation to head of household), their reason for moving into the household, and their reported time spent living in the community. 
\end{itemize}

We will also evaluate the following metrics  quantifying HIV status and HIV RNA suppression status among mobile populations.
\begin{itemize}
\item  HIV prevalence among in-migrants
\item The proportion of HIV-positive adult residents at year 3 who are in-migrants
\item The proportion of all  HIV-positive adult residents with (i) HIV
 RNA$\geq$500 copies/ml  and (ii) HIV RNA $>$100,000 copies/ml at year 3 who are in-migrants
\item Among HIV-positive adult residents at year 3 who are  in-migrants, the proportion with (i) HIV
 RNA$\geq$500 copies/ml and (ii) HIV RNA $>$100,000 copies/ml at year 3
\item The proportion of baseline HIV-positive  adults who out-migrate  by year 3
\item Among all adults who are known to be HIV-positive at year 3, are classified as out-migrants, and have measured HIV RNA, the proportion with (i) HIV RNA$<$500 copies/ml at year 3, (ii) HIV RNA $>$100,000 copies/ml at year 3 -  overall and stratified by FUY3 testing location.
\end{itemize}



\subsection{Descriptive and community-level explanatory analyses}

The community-specific  cumulative incidence (as estimated  for the primary outcome) and the community-specific incidence rates (incident cases per 100 person-years at risk) over the three  year period will be reported. In estimating incidence rates, person-years at risk for individuals who test HIV-negative at both baseline and year 3 will be calculated from date of baseline negative test to date of year 3 negative test; person-years at risk for incident HIV infections will be calculated from date of baseline negative test to midway between date of baseline HIV-negative test and year 3 HIV-positive test.

We will consider the following independent explanatory variables, estimated as specified in the prior section:

		\begin{itemize}
		\item  HIV prevalence (at baseline and using the average of baseline and year 3). 
		\item HIV RNA metrics: total unsuppressed person-time/total adult person-time; proportion of HIV-positive adults with HIV RNA$>500$ copies/ml and with $>$100,000 copies/ml, at baseline and using the average of baseline and year 3; and proportion of total adult population with HIV RNA$>500$ copies/ml and with $>$100,000 copies/ml, at baseline and using the average of baseline and year 3.
		\item Male circumcision coverage, 
		using the average coverage at baseline and year 3. 
		\item Mobility: proportion of the adult population who are in-migrants at year 3,  proportion of the baseline population who have out-migrated by year 3, proportion of all adults and HIV-positive adults who are unsuppressed in-migrants.
			\end{itemize}
			 We will conduct these analyses overall, and stratified by sex.  In sex-stratified analyses, we will consider as additional predictors corresponding to metrics among the opposite sex.

Unadjusted and adjusted associations between each explanatory variable and  HIV incidence will be evaluated. Community-level Poisson regression of the community-specific HIV incidence on the community-level explanatory variables, in turn and jointly, will be used to estimate the unadjusted and adjusted (conditional) relative risks and rates. 
 We will also conduct analyses under the additional assumption that individual outcomes are independent given the explanatory variables and additional individual-level covariates included in the adjustment set (listed below). 
 Individual-level Poisson regression will be used to estimate the relative risk/rate of incident HIV infection per unit change in each explanatory variable, adjusted for the remaining explanatory variables, region, and additional individual-level  risk factors measured at study baseline (listed below).  Inference will employ robust standard error estimates based on an exchangeable working covariance matrix. 
   In addition to the explanatory variables above and region, these individual-level analyses will adjust for the following baseline individual-level covariates: 
 sex, age, marital status, education, occupation, household wealth index, mobility, 
 alcohol use, contraceptive use, 
 polygamy, and self-reported prior HIV testing. 




\FloatBarrier

\section{Intervention uptake: Testing, HIV care cascade, and population-level HIV RNA metrics} \label{uptake}

The study intervention aims to achieve high annual levels of HIV testing coverage, rapidly initiate ART among all HIV+ individuals, and retain these individuals in care with HIV viral suppression (plasma HIV RNA$<$500 copies/ml). The control arm of the study also aims to achieve high levels of HIV testing coverage at study baseline, and provides a clinical officer to support adherence to  ART  guidelines, which evolved  over the course of the study. To evaluate intervention uptake in both study arms over time, with implications for understanding both health of HIV-positive individuals and HIV transmission potential, we will conduct the following analyses. Throughout, we will define prior HIV diagnosis and ART initiation (i) using Ministry of Health records only, and (ii) incorporating self-report.

\subsection{Testing uptake in intervention and control arms}

We will provide descriptive statistics to characterize testing coverage and other services by treatment arm, by community, and over time among the open cohort of adult ($\geq$15 years at year $t$, $t \in \{0,1,2,3\}$) community residents. Community residence in a given year will be determined by the baseline enumeration combined with the re-census of the adult population at year 3. An individual will be considered resident in the community at year $t$ if he or she was resident at baseline and is not known to have died or out-migrated by year $t$, or if he or she was not a resident at baseline and is reported to have in-migrated by year $t$. 
In this open cohort of adult residents, we will report for each year $t$: (i) the proportion of the population with known HIV status at the close of year $t$ testing, and (ii) the proportion of the population ever tested for HIV by the close of year $t$ testing. Known HIV status at year $t$ will be defined as  either having a prior HIV diagnosis or having no prior HIV diagnosis and a positive or negative HIV test result at year $t$ testing. 
Ever testing for HIV by year $t$ will be defined as having a prior HIV diagnosis or any prior HIV rapid test result by close of year $t$ testing. Pre-baseline and secondary analyses will also incorporate self-report of prior HIV testing and self-reported prior test results.     

For each year, by arm and by region, we will report the portion of the eligible population (i) attending the CHC, (ii) seen at tracking, and (iii) contacted at either CHC or tracking. We will further report  the proportion of the eligible population receiving specific screening and testing services. These include HIV testing (among those not already known to be HIV-positive), plasma HIV RNA level testing (among HIV-positive individuals), and hypertension and diabetes screening.  We will characterize predictors of not attending the CHC and of not being contacted at either the CHC or tracking. We will also report the extent and components of community mobilization carried out prior to testing campaigns.

We will describe clinic-specific timing of the uptake of each of the components of streamlined HIV care: rapid ART start, appointment reminders, viral load counseling, and tiered tracking for missed visits. We will also provide clinic-specific timing of the implementation of evolving national ART guidelines in the control arm. Further analyses of  linkage, retention, and suppression over time in intervention and control communities are described in detail below. 

\subsection{Cross-sectional cascade coverage in open cohort of HIV-positive individuals}\label{cascade.open}

 At baseline and year 3 ($t=\{0,3\}$), we will estimate the proportion of all HIV+ adults who are previously diagnosed, the proportion of previously diagnosed adults who have ever initiated treatment, the proportion of those ever on ART who are currently suppressed, and overall population-level suppression (the proportion of all HIV+ who are currently suppressed). 
 %
 This analysis will be based on the open cohort of individuals who are aged $\geq$15 years and are community residents at $t$, as defined above. Primary analyses will include baseline residents (regardless of stability) and in-migrants identified at follow-up year 3; secondary analyses will restrict to baseline stable residents. We will also report simple descriptive analyses of HIV status and suppression among non-residents with measured HIV RNA levels.
 We will use TMLE to adjust for potentially differential measurement of HIV status and viral loads, using methods detailed in \citep{PetersenCascade2017, Balzer2017CascadeMethods};  unadjusted numbers and proportions will also be reported. We will conduct sensitivity analyses to incorporate self-report of prior diagnoses and ART use and to adjust for potentially differential measure of prior diagnoses and ART use. 

We will report cascade and suppression estimates by community, by treatment arm, and for the following  strata: region, sex, age (15-24 years; 25+ years), mobility ($<1$mo away; $\ge$1mo away), students, and fishermen. We will also generate annual estimates in the intervention arm.
 
 We will estimate the effect of the randomized intervention on population-level suppression at year 3 and test the null hypothesis that  population-level viral suppression at year 3 is the same between the intervention and control arms. To test this null hypothesis, we will use methods analogous to those used for the primary study endpoint. Specifically, we will estimate a community-specific outcome (population-level viral suppression at year 3) and then estimate the effect with a community-level TMLE using data-adaptive selection of adjustment variables from a pre-specified set. For this outcome, the pre-specified candidate adjustment variables are the proportion of the baseline HIV-positive population with HIV RNA$<500$ copies/ml, and the proportion of baseline HIV-positive adult population below the age of 25. Our primary analysis will give equal weight to individuals (weight communities relative to the size of their HIV-positive population); secondary analysis will give equal weight to communities. Subgroup analyses will include region (using an unadjusted stage II estimator), sex, age,  mobility, and circumcision status. 
 
Within each treatment arm, we will also estimate the change in population-level suppression between baseline and year 3 and test the null hypothesis that population-level viral suppression remained constant between baseline and year 3.   First, we will estimate  change in population-level suppression between baseline and year 3, adjusting for missing HIV serostatus and  HIV RNA measures, but without adjusting for changes in the HIV-positive population over time. This estimate will be based on a comparison of the time point-specific TMLE-based population suppression estimates described above. 
Second, we will estimate the change in population-level suppression over time adjusted for any changes in the distribution of individual-level characteristics among those with known HIV and viral suppression status over time. This parameter,  in addition to accounting for changes in measurement patterns, adjusts for any changes in the distribution of the HIV-positive population over time. 
Estimation will be based on a TMLE that pools over individuals and years, with time as the exposure of interest.  Standard errors will be estimated based on the estimated influence curve, treating the individuals as independent within communities.
In both approaches, community will be adjusted for as a fixed effect. Outcome and propensity score models will  be fit data-adaptively using Super Learning. Adjustment variables will include age, sex, occupation, education, mobility, wealth, marital status, and testing location.   

For the same primary and secondary populations, 
we will use analogous methods to estimate the proportion of all HIV-positive adults in the following viral load strata at baseline and year 3: $<$1000 cps/ml, 1000-100,000 cps/ml, and $>$100,000 cps/ml.
We will estimate the effect of the randomized intervention on  (i) the population-level proportion with viral loads $>$100,000 cps/ml, and (ii) the population-level proportion with viral loads $\geq$1000 cps/ml.
Using methods analogous to those used for suppression, we will estimate the ratio of the mean of these community-level outcomes between treatment and control arms and the change in their mean between baseline and year 3, and test the corresponding null hypotheses. 

\subsection{Longitudinal HIV RNA levels in the closed cohort of baseline HIV-positive individuals}\label{cascade.closed}

For the subgroup of adult (aged $\geq$15) baseline residents diagnosed with HIV at or before baseline, we will estimate the proportion virally suppressed at year 3 by treatment arm.  
We will censor at death and out-migration, 
using TMLE to adjust for potentially non-differential measurement of viral loads and for censoring, as  detailed in \citep{PetersenCascade2017, Balzer2017CascadeMethods};  unadjusted proportions will be reported as secondary analyses. The primary analysis will include all baseline residents; secondary analyses will restrict to baseline stable residents.

We will report suppression estimates by region, by community, by treatment arm, and for the following baseline strata:  cascade subgroup (no prior HIV care, prior care but no prior ART initiation, prior ART but not suppressed, and suppressed),  CD4+ T cell count ($<$350 cells/$\mu$l, 350-500 cells/$\mu$l, $>$500 cells/$\mu$l) region,  sex,   age (15-24 years; 25+ years),  mobility ($<1$mo away; $\ge$1mo away), students, and fisherman. Within the intervention arm, we will report analogous estimate for years 1 and 2.

 We will estimate the effect of the randomized intervention on the proportion of the baseline HIV-positive population who are suppressed at year 3 and test the null hypothesis of no difference in proportion suppressed between intervention and control, using the same approach as for the open cohort of HIV-positive adults (Section \ref{cascade.open}). 

Using analogous methods, we will  estimate the proportion of baseline HIV-positive individuals in the following viral load strata at year 3: $<$1000 cps/ml, 1000-100,000 cps/ml, and $>$100,000 cps/ml. We will estimate the effect of the randomized intervention on the  proportion of baseline HIV-positive individuals with viral loads $>$100,000 cps/ml and test the corresponding null hypothesis.

In a complimentary analysis of the closed cohort of baseline HIV-positive individuals, we will report the probability of being in each of the following mutually exclusive and exhaustive states  at baseline and follow-up year 3: died, migrated out of the community, 
newly diagnosed, previously diagnosed but never on ART, had initiated ART but are unsuppressed, and currently suppressed. This  analysis will adjust for informative viral load measurement, using methods described in \cite{PetersenCascade2017, Balzer2017CascadeMethods}; unadjusted estimates will also be reported.

We will evaluate baseline predictors of  (i) having a plasma HIV RNA level $>$500 copies/ml at year 3, and (ii) having a plasma HIV RNA $>$100,000 copies/ml at year 3.
 Specifically, we will report unadjusted associations and adjusted variable importance measures on the relative scale (statistical analogs of the causal risk ratio), treating each baseline predictor in turn as the intervention variable, and the remainder as the adjustment set. Variable importance measures will be estimated with pooled individual-level TMLE, adjusted for community as a fixed effect, and treating the household as the independent unit for variance estimation. Using methods described in \cite{PetersenCascade2017, Balzer2017CascadeMethods}, we will further adjust for potentially informative censoring and missing viral load measures.
Baseline predictors considered will include region, sex, age, marital status, education, occupation, household wealth index,  mobility,  alcohol use, and testing location. 
%

\subsection{Time to linkage and ART initiation}\label{art.start}


For the subgroup of adult residents diagnosed with HIV but not yet on ART at baseline, we will conduct longitudinal analyses to estimate the probability of initiating ART. 
Specifically, among adult residents testing HIV-positive at either the baseline CHC or  tracking, and not currently in care, we will evaluate the probability of ART initiation over time. 
Time-zero will be the date of the baseline campaign or home-based contact. Primary analysis will right-censor at the time of death, out-migration, 
 or close of community-specific follow-up year 3 testing.
  Secondary analyses will (i) not censor at out-migration,
  and  (ii) censor at out-migration, but treat death before ART initiation as a failure to initiate. 
The primary analysis will include all baseline residents; secondary analyses will restrict to baseline stable residents.
  
 In estimating longitudinal probabilities of ART initiation, 
 we will use Kaplan-Meier analyses, 
 and will plot the corresponding survival curves by treatment arm and by region. 
Using a two stage approach, we will estimate the effect of the randomized intervention on probability of  initiating ART  by 6, 12, and 24 months. We will test the corresponding null hypotheses of no intervention effect, using a community-level TMLE analogous to that used to compare viral suppression between arms, weighting individuals equally and as candidate adjustment variables  the proportion of  adult residents known to be HIV-positive at baseline but not yet on ART at baseline who are (i) aged 15-24 years, and 
(ii) new HIV diagnoses.

We will report estimates  by treatment arm, by region,  by community,
 and for the following baseline strata: no prior HIV care at at baseline, prior HIV care without a record of prior ART at baseline,   CD4 count ($<$350, 350-500, $\geq$500), sex, age (15-24 years; 25+ years), and baseline mobility ($<$1mo away;  $\geq$1mo away). 
Among individuals who initiate ART in the control communities, we will document reason for ART start.
We will provide descriptive statistics of the following variables at the time of ART initiation: 
CD4 count, sex, age, occupation, and mobility ($<1$mo away; $\ge$1mo away) by treatment arm and by community. 

We will evaluate baseline predictors of not initiating ART within 12 months. 
 Specifically, we will report unadjusted associations and adjusted variable importance measures on the relative scale (statistical analogs of the causal risk ratio), treating each baseline predictor in turn as the intervention variable, and the remainder as the adjustment set. Variable importance measures will be estimated with pooled individual-level TMLE, adjusted for community as a fixed effect, and treating the individual as the independent unit for variance estimation.
 Baseline predictors considered will include region, sex, age, marital status, education, occupation, household wealth index,  mobility,  alcohol use, and testing location, and CD4 count.


\subsubsection{Time to Linkage}

Time to ART initiation is a function of both time to linkage and time from linkage to ART start, and both may be differentially impacted by the study intervention.  To further disaggregate these steps in the care cascade, we will conduct  analyses analogous to the above, among individuals who are not currently in HIV care at baseline (either newly diagnosed with HIV at baseline, or previously diagnosed but not currently in HIV care),  using the alternative outcome of linkage to care, defined as the first recorded visit to an HIV clinic following baseline. 

\subsection{Retention in HIV Care}

Among HIV-positive adult residents who initiate ART during study follow-up (between the start of baseline community-specific CHC and start of follow-up year 3  community-specific testing),
we will conduct longitudinal analyses to estimate the probability of being retained in HIV care over time. Retention failure is defined as more than 90 days late to a scheduled 12-month follow-up  \citep{Brown2016retention}. 
Retention failures include those living in the community but not engaged in care, those who move out of the community without a documented transfer,  and those otherwise lost to follow-up. 
%
Time-zero will be the date of ART initiation.  The primary analysis will right-censor at time of death, 
documented transfer to a non-SEARCH clinic, or  close of community-specific follow-up year 3 testing. Secondary analyses will also censor at date of outmigration (even if no transfer documented). The primary analysis will include all baseline residents; secondary analyses will restrict to baseline stable residents.

In estimating longitudinal probabilities of retention over time, we will conduct both unadjusted (Kaplan-Meier) analyses and corresponding analyses adjusted for potentially informative censoring using TMLE. We will plot the resulting survival curves by treatment arm and by region. 
We will estimate the effect of the randomized intervention on the probability of being retained in care 12 months  after ART initiation  and test the corresponding null hypothesis of no intervention effect, using a community-level TMLE analogous to that used to compare time to ART initiation between arms -  weighting individuals equally and with candidate adjustment variables consisting of the  proportion of adults newly diagnosed with HIV at baseline who are aged $<$25 years and the proportion of adults newly diagnosed with HIV at baseline who are mobile ($\geq$1 month of past year spent outside the community). 



Any differences observed in retention between treatment arms may be attributable in part to differences in the population initiating ART (the population of ART initiators evaluated may differ by study arm in terms of how challenging they are to retain in care). We will, therefore,  conduct complimentary analyses to 
estimate the probability of  having initiated ART and remaining retained in HIV care.
This outcome captures the total intervention effect on both ART start and subsequent retention among starters.

We will report estimates by treatment arm, by region, by community, and for the following strata: no prior HIV care at baseline, prior HIV care  without a record of prior ART at baseline, study year of ART initiation,  CD4 $<$500 versus $\geq$500 at time of ART initiation, and $\leq$30 days versus $>$30 days between date of first HIV-positive test during SEARCH follow-up and ART initiation.


We will provide descriptive statistics as well as  evaluate unadjusted and adjusted associations between baseline individual-level characteristics and non-retention.
 Specifically, we will report unadjusted associations and adjusted variable importance measures on the relative scale (statistical analogs of the causal risk ratio), treating each baseline predictor in turn as the intervention variable, and the remainder as the adjustment set. Variable importance measures will be estimated with pooled individual-level TMLE, adjusted for community as a fixed effect, and treating the individual as the independent unit for variance estimation.
  Baseline predictors considered will include region, sex, age, marital status, education, occupation, household wealth index,  mobility,  alcohol use,  testing location, 
CD4 count at ART start ($<$500; $\geq$ 500), and time from diagnosis to ART start ($\leq$ 30 days; $>$ 30days).

We will implement analogous analyses to evaluate retention among  individuals with a prior history of ART use at study baseline who have at least one documented clinic visit post-baseline. For this subgroup, time-zero is the date of first post-baseline clinic visit, and the primary outcome is, as above, retention 12 months after this date.


\subsection{Analysis of interim testing}


To further understand the impact of annual population-based testing, for communities in the intervention arm we will report the following descriptive statistics, overall and by region:
\begin{itemize}
\item Number of newly diagnosed HIV-positive individuals seen during population-based testing at time $t$; their demographics (sex,  age (15-24 years; 25+ years), and  baseline mobility ($<1$mo away; $\ge$1mo away)), CD4+ T cell count, HIV RNA level,  prior SEARCH testing history, and residence status  (baseline stable resident, baseline non-stable resident, in-migrant, non-resident); and the proportion of all HIV-positive individuals seen  at time $t$ who are newly diagnosed, overall and stratified by prior HIV testing history.   
\item Number of previously diagnosed HIV-positive individuals with no history of ART seen during population-based testing at time $t$; their demographics (sex,  age (15-24 years; 25+ years), and  baseline mobility ($<1$mo away; $\ge$1mo away)), CD4 count, HIV RNA level,  prior SEARCH testing history, and residence status (baseline stable resident, baseline non-stable resident, in-migrant, non-resident); and the proportion of all HIV-positive individuals seen at time $t$ who are previously diagnosed but with no prior history of ART. 
\item Among new diagnoses, time to linkage, time to ART initiation, and the proportion suppressed one  (for new diagnoses at year 1 and year 2) and two years following diagnosis (for new diagnoses at year 1) (Noting that analysis of these outcomes among individuals testing baseline HIV-positive at baseline are described above). 
\end{itemize}

Analogous analyses will be conducted restricting to incident HIV infections identified at interim campaigns (i.e. restricting new diagnoses to those with a prior negative HIV test). To quantify background (non-SEARCH) diagnosis and linkage rates, we will further report in both arms the  number and proportion of new diagnoses and of incident infections during the three years of study follow-up using the earliest record of HIV-positive status recorded in Ministry of Health clinical records or linkage to the tuberculosis registry rather than SEARCH annual population-based testing. 

\section{Community adult health outcomes }\label{health}


A primary aim of the SEARCH Study is to understand the intervention's effect on the health of the overall community as well as the health of people living with HIV. 
This section describes evaluation of the adult health outcomes of mortality, tuberculosis (TB), and non-communicable diseases (NCDs), specifically diabetes (DM) and hypertension (HTN). Descriptive analyses of ART toxicity and resistance are also given.

\subsection{Mortality}
The SEARCH intervention may reduce mortality by enabling earlier diagnosis, earlier ART initiation, and improved ART retention and suppression outcomes among HIV-positive individuals, as well as, to a lesser degree, reducing exposure to infectious outcomes like TB.  In addition, the comprehensive mortality data collected as part of SEARCH provides an opportunity to accurately quantify overall mortality in this rural East Africa setting in the context of universal test and treat, as well as community-wide HIV testing and linkage to care at study baseline (the active control arm of the trial). In this section, we specify analyses to compare mortality among baseline HIV-positive individuals and among the overall adult population between intervention and control arms, as well as additional descriptive and explanatory analyses.

\subsubsection{Mortality among HIV-positive adults}
Mortality among HIV-positive adults will be compared between arms using a two-stage approach analogous to that employed for the HIV incidence, ART initiation, and plasma HIV RNA suppression outcomes. The first stage will estimate community-specific mortality risk among adults known to be HIV-positive at or before study baseline, both overall and restricted to those with no record of ART use prior to baseline. Kaplan-Meier estimators will be used to estimate post-baseline survival in each community. In primary analyses, failure will be defined as death due to illness, and follow-up time will be censored at death to other causes or out-migration from the community; in secondary analyses, failure will be defined as death due to any cause. Primary analyses will estimate survival among baseline stable adult residents; sensitivity analyses will include baseline non-stable residents. In the second stage, risk of mortality by three years (or, if less than three years, by the minimum Phase 1 follow-up time across communities) will be compared between arms, weighting individuals equally, and with candidate adjustment variables consisting of the proportions of the analytic population with baseline CD4+ T cell count $\leq$50 cells/$\mu$l and with baseline CD4+ T cell count $\leq$350 cells/$\mu$l, as well as for analyses of all baseline HIV-positive adults (including those on ART at baseline) the proportion of the analytic population with baseline plasma HIV RNA level $<$500 copies/ml. 

To distinguish between the impact of streamlined linkage and ART delivery in the intervention arm from the impact due to differences in CD4+ T cell count eligibility threshold for ART initiation at study baseline, we will further conduct subgroup analyses testing the null hypothesis of no difference in mortality risk between arms within subgroups defined by baseline CD4+ T cell count: $\leq$350 cells/$\mu$l and $>$350 cells/$\mu$l. The intervention effect will be estimated using a two stage approach identical to the primary analysis described above, with the exception of the following modification to the candidate CD4-based adjustment variables used in Stage 2:
\begin{packed_item}
\item For the subgroup of baseline CD4+ T cell count $\leq$350 cells/$\mu$: the proportion with CD4+ T cell count $\leq$100 cells/$\mu$l
\item For the subgroup of baseline CD4+ T cell count $>$350 cells/$\mu$: the proportion with CD4+ T cell count 350-500 cells/$\mu$l
\end{packed_item}

We will also report the following  descriptive analyses of mortality among baseline HIV-positive individuals:
\begin{packed_item}
\item Kaplan-Meier-based survival estimates, stratified by baseline cascade status (no 
prior
HIV care, prior HIV care but no prior ART, prior ART but with plasma HIV RNA level $\geq$500 copies/ml at baseline, and plasma HIV RNA level $<$500 copies/ml at baseline)
\item Estimated mortality rates (per 100,000 person years) with person-time at risk beginning at the start of Phase I and ending at death, out-migration, or end of phase 1. Mortality rates will be reported among all baseline HIV-positive individuals, and among baseline HIV-positive adults aged 15-59.
\end{packed_item}
Primary analyses will be based on baseline stable residents; secondary analyses will include non-stable residents.

We will further estimate unadjusted and adjusted predictors of death among the baseline HIV-positive population. A pooled individual-level TMLE including community as a fixed effect will be used to estimate adjusted and unadjusted variable importance measures (on the relative scale) for the following covariates: sex, age, marital status (including widow as a separate category if sufficient data support exists), education, occupation, household wealth index, mobility, having a baseline HIV+ adult (other than the current individual) in the same household, baseline CD4+ T cell count, and baseline plasma HIV RNA level $<$500 copies/ml. 

Finally, we will estimate the mortality rate due to illness, and due to any cause, following ART initiation. Person-time at risk will begin at the time of ART initiation and end at the first of death, out-migration or end of Phase 1. We will compare this rate between arms, using a two stage TMLE, weighting person-time equally and with candidate adjustment variables consisting of the proportions of baseline HIV-positive individuals not on ART at baseline with CD4+ T cell count $\leq$50 cells/$\mu$l and CD4+ T cell count $\leq$350 cells/$\mu$l at baseline. In interpreting this cross arm comparison, we note that any difference between arms will be a function of both any intervention effect on the underlying mortality risk of individuals initiating ART (i.e. via intervention effects on time to ART initiation overall and within subgroups), as well as any effect of the intervention on survival post-ART initiation. In descriptive analyses, Kaplan-Meier-based survival following ART initiation will also be estimated.

\subsubsection{Mortality among all adults}
Mortality rate (all-cause, and due to illness) will be estimated among all baseline stable residents (primary) and among all baseline residents (regardless of stability; secondary). Person-time at risk will begin at the first of the start of Phase 1, or age $\geq$15 years, and will end at the first of death, out-migration, or the end of Phase 1.  We will compare this rate between arms, using a two stage TMLE weighting person-time equally, with candidate adjustment variables consisting of baseline HIV prevalence and the proportion of adult residents falling in lowest quintile of household socioeconomic index.

Additional descriptive analyses of mortality in the adult population will include:
\begin{packed_item}
\item A description of the distribution of causes of death (illness, childbirth, homicide, accident, suicide) among adult deaths, stratified by arm.
\item Age-adjusted mortality rates, using direct standardization to the WHO standard age distribution, and stratified by arm and by baseline HIV status.
\item Unadjusted and adjusted predictors of mortality, analogous to the variable importance measures estimated for the baseline HIV-positive population.
\end{packed_item}

\subsubsection{Comparison of mortality between HIV-positive and  HIV-negative adults}
We will compare mortality rates between the adult baseline HIV-positive and baseline HIV-negative populations, over all communities, and stratified by region and treatment arm. Mortality rates will be standardized to the pooled age-sex distribution of the populations being compared. We will compare survival curves over time between the adult baseline HIV-positive and HIV-negative populations, using both unadjusted Kaplan-Meier estimators, and TMLE of baseline HIV-status-specific survival curves, adjusted for baseline predictors of mortality  including age, and including community as a fixed effect.

\subsection{Tuberculosis (TB)} 
More rapid initiation and effective ART delivery as a result of streamlined care may reduce the incidence of TB disease and death among HIV-positive individuals by  reducing susceptibility to TB disease among HIV-positive persons, and by reducing death due to HIV-associated TB. In this section we describe analyses to compare the composite outcome of HIV-TB and death between treatment arms, as well as additional descriptive analyses of incident active TB and TB-associated morbidity.

\subsubsection{Incident Active HIV-associated Tuberculosis}
Our primary HIV-TB outcome for comparison between intervention and control arms will be estimated among baseline stable adult residents who either test HIV-positive at baseline or who have a missing baseline HIV test. Our motivation for including individuals with a missing baseline HIV test is that individuals who are harder to access for HIV testing may be those individuals at higher risk of HIV-TB. In sensitivity analyses we will (i) include non-stable residents; and (ii) restrict to individuals known to be HIV-positive at baseline. The population will exclude individuals with an active TB diagnosis within 6 months prior to the start of Phase 1. In this population, we will estimate risk of the composite outcome of (i) death due to illness or (ii) incident active TB disease with an HIV diagnosis recorded at or prior to  date of TB diagnosis. The primary HIV-TB analysis is focused on this composite outcome 
based on evidence that a substantial portion of death due to illness among HIV-positive persons is due to undiagnosed TB.  

Comparison of  HIV-TB outcomes between arms will be based on a two stage analysis. In the first stage, community-specific Kaplan-Meier estimators will be used to estimate the risk of the composite outcome by three years (or the minimal time for which all communities have follow-up), censoring at death due to other causes or out-migration.  In the second stage, these community-level risk estimates will be compared between arms using TMLE, weighting individuals equally, and with a candidate adjustment set consisting of (i) baseline HIV prevalence among adults, and (ii) the number of TB cases diagnosed in year prior to baseline divided by the number of baseline adult residents. 

In secondary analyses, we will evaluate the non-composite outcome of HIV-associated TB (censoring at outmigration or death due to any cause) among the full adult population (irrespective of baseline HIV status). Comparison of these secondary outcomes between arms will be implemented analogously to the primary TB outcome, using community-specific Kaplan Meier estimators to estimate 
risks, and comparing estimated risks between arms using TMLE, weighting individuals equally, and with the same candidate adjustment set as for the primary TB outcome. 

The above analyses will also be implemented stratifying on region (unadjusted Stage 2 estimator only) and for the following subgroups: (i) baseline HIV-positive adults with baseline CD4+ T cell count $\leq$500 cells/$\mu$l; (ii)  baseline HIV-positive adults with baseline CD4+ T cell count  $>$500 cells/$\mu$l;  (iii) baseline HIV-positive adults with baseline plasma HIV RNA level $\geq$500 copies/ml; and (iv) men and women.

\subsubsection{Additional TB analyses}
We will implement the following additional secondary analyses to characterize incident active TB by HIV status, to determine the predictors of HIV-associated active TB, and to compare the clinical outcomes of participants with active TB in intervention vs. control communities. The overall goal of these secondary analyses is to better understand potential changes in the epidemiology of HIV-associated TB, including how the SEARCH  intervention might impact the risk of developing active TB and  clinical outcomes.
\begin{itemize}
\item We will report Kaplan-Meier survival curves for the time-to-event outcomes corresponding to the primary and secondary HIV-TB outcomes, by region and arm, and within the following subgroups: baseline HIV-positive adults with CD4+ T cell count $\leq$500 and $>$500 cells/$\mu$l at baseline; baseline HIV-positive adults with plasma HIV RNA level $\geq$500 and $<$500 copies/ml at baseline; men and women;  youth (aged 15-24 years) and older individuals (aged $\geq$ 25 years).
\item We will use Kaplan-Meier estimators to evaluate the risk over time for developing (i) TB, and (ii) HIV-TB among (i) the baseline HIV-negative population, and (ii) the baseline HIV-positive population, censoring at death or outmigration. We will report the corresponding survival curves.
\item We will estimate annualized TB and HIV-TB incidence rates in an open cohort of adult residents, overall and stratified by baseline HIV status. Person-time at risk will begin at the first date at which the individual is a community resident (either a baseline resident, or has in-migrated, using in-migration date ascertained at year 3) and aged $\geq$15 years, and will end at the first of diagnosis of active TB, death, outmigration, or end of Phase 1. Annual TB incidence rates will be reported by community and by treatment arm (overall and within region).
\end{itemize}

\subsubsection{Predictors of incident TB}
We will report unadjusted associations and adjusted variable importance measures on the relative scale (statistical analogs of the causal risk ratio), treating each baseline predictor in turn as the intervention variable, and the remainder as the adjustment set. Variable importance measures will be estimated with pooled individual-level TMLE, using the community as the independent unit when estimating variance, and with inference based on the t-distribution. The population, outcome, and right-censoring variables will be defined as for the primary outcome; analogous to predictors of HIV seroconversion, secondary analysis will consider community as a fixed effect. Baseline predictors considered will include: sex, age, education, household wealth, mobility, alcohol use, and among baseline HIV+, baseline CD4+ T cell count and plasma HIV RNA level. Predictors will be evaluated overall and stratified by baseline HIV status. 

\subsubsection{Characterization of active TB cases by HIV status}
We will conduct the following analyses to evaluate how individuals diagnosed with active TB and the clinical presentation of active TB varies over time and by HIV status.
Specifically, we will describe demographic and clinical characteristics of incident active TB cases diagnosed during Phase 1, overall and stratified by (i) HIV status, (ii) intervention vs. control arm, and (iii) by study year. We will provide descriptive statistics of the following characteristics of TB cases: demographics (age, sex, wealth index, education); mobility; TB disease site (pulmonary, extra-pulmonary); AFB smear; TB disease type (new, re-treatment, failure, default); and, among HIV-positive individuals, CD4+ T cell count and plasma HIV RNA level at time of TB treatment start. 

We will also calculate the empirical proportion of  incident TB cases that are HIV-associated (defined, as for the primary outcome, as having an HIV diagnosis at or prior to TB diagnosis date), stratified by intervention arm and by year. We will compare this proportion between intervention and control arms using the two stage approach used for the primary TB outcome.

\subsubsection{Clinical Outcomes among HIV-associated TB cases}
We hypothesize that TB clinical outcomes following TB treatment start will be positively impacted by the SEARCH intervention due to earlier diagnosis of HIV, universal access to ART, and streamlined HIV care delivery. These impacts may include reduced mortality during TB treatment, reduced risk of IRIS (defined as CD4$<$100 at TB treatment start), and more rapid time to ART start (if not on ART at TB diagnosis) in intervention communities compared to control.

Among individuals diagnosed with HIV-associated, incident active TB disease in Phase 1, we will estimate the mortality rate due to illness during TB treatment, with person-time at risk beginning on date of TB treatment initiation and ending on first of out-migration, death due to other causes, TB treatment end date, or end of Phase 1. We will  compare this rate between intervention arms with a two-stage approach analogous to the approach used for the primary TB outcome, weighting person-time equally, and with candidate adjustment variables consisting of proportion of baseline HIV-positive individuals with CD4+ T cell count$\leq$350 cells/$\mu$l, and proportion of baseline HIV-positive individuals with HIV RNA level $<$500 copies/ml.
As with mortality rates following ART initiation, we note any intervention effect may be mediated in part by impacts on the characteristics of TB treatment initiators at time of TB treatment start.  

We will also report the following descriptive analyses:
\begin{packed_item}
\item The proportion of incident active TB cases among HIV-positive individuals with CD4+ T cell count $<$200 cells/$\mu$l that develop IRIS
\item The proportion of incident active TB cases with prior ART use at time of TB diagnosis
\item Time to ART initiation following incident active TB diagnosis among individuals who are HIV-positive and not on ART at the time of TB diagnosis
\end{packed_item}


\subsection{Non-Communicable Diseases}\label{Sec:NCD}

At baseline, all communities received population-based hypertension (HT) screening and referral for treatment according to national guidelines. Intervention communities further received annual population-based HT screening, with HT treatment for HIV-positive and HIV-negative individuals delivered using an integrated streamlined-care delivery model. Screening and care for diabetes (DM) was delivered analogously in the intervention and control arms, with the exception that DM screening at baseline was limited to Ugandan communities. The SEARCH intervention may thus improve HT and DM control through earlier diagnosis and through more effective treatment delivery. As with HIV RNA suppression and HIV incidence, baseline population-based testing in the control arm may also improve HT/DM control over time. In this section, we describe analyses to evaluate these hypotheses among adults aged $\geq$ 30 years at follow-up year 3. As throughout, to ensure comparable data structures, analyses comparing intervention and control arms only make use of data collected at population-based testing at baseline and at follow-up year 3.

We define HT control  as having  at least one systolic blood pressure (BP) measurement $<$140 mmHg  and  at least one diastolic BP measurement $<$90 mmHg.
 In other words, uncontrolled HT is defined as all  systolic BP measures $\geq$140 mmHg or all  diastolic BP measures $\geq$ 90 mmHg (requiring that at least one BP measure was recorded).
We define prevalent HT as current or previous (i) self-report of a prior diagnosis, or (ii) uncontrolled blood pressure.
 For HIV-positive and HTN prevalent persons, dual-control is defined  as joint control of HT and suppressed viral replication ($<$500 copies/ml).  
 The metrics for DM are defined analogously, with DM control defined as a finger-prick blood glucose $\leq$11 mmol/L and prevalent DM defined as previous or current (i) self-report of a prior diagnosis, or (ii) uncontrolled blood glucose. 

\subsubsection{Hypertension (HT) control among individuals with prevalent HT at year 3}\label{HTN3}


HT control among adults aged $\geq$ 30 years and with prevalent HT at year 3 will be compared between arms using the two-stage approach, detailed above.
The first stage will estimate the community-specific proportion of adults with prevalent HT at year 3 who have their HT controlled at follow-up year 3, both overall and among those known to be HIV-positive at year 3. 
Using an approach analogous to that used to estimate population-level viral suppression, these population-level proportions will be estimated for each community using an individual-level TMLE, adjusting for incomplete measures of both HT disease status at year 3 and incomplete measures of disease control at year 3. Secondary analyses will restrict to individuals known to have HT at year 3, adjusting for incomplete measures of disease control.
Among adults  with prevalent HT and HIV at year 3, we will also estimate the community-specific proportion with dual-control (both controlled HT and plasma HIV RNA level $<$500 copies/ml) at follow-up year 3, adjusting for missing measures of both HT control and plasma HIV RNA at year 3.

 Primary analyses will be restricted to baseline stable residents; sensitivity analyses will include baseline non-stable residents and in-migrants identified in year 3 testing.
Primary analyses will condition on being alive and resident in the community at follow-up year 3; secondary analyses will further adjust for censoring by death and out-migration.
Sensitivity analyses will be based on unadjusted empirical proportions.

In the second stage, community-specific estimates of control at follow-up year 3 will be compared between arms, weighting individuals equally, and with the following candidate adjustment variables: 
\begin{packed_item}
\item baseline CHC testing coverage and baseline prevalence of having a body mass index (BMI) $>$ 24 when estimating the effect on
HT control at follow-up year 3
\item baseline CHC testing coverage and baseline dual-control when estimating the effect on dual-control at follow-up year 3 
\end{packed_item}
We will also test the null hypothesis of no intervention effect stratifying on region (for each population) in unadjusted analyses.
Using the analogous two-stage approach, we will also report estimates of HT control and HIV-HT dual-control stratified by community, by intervention arm, and within intervention by region, sex, and
age (30-44 years, 45-59 years, 60+ years).



\subsubsection{HT control among individuals with prevalent HT at baseline}\label{HTN0}

HT control among adults ($\geq$30 years) known to have HT at baseline (via self-report or elevated blood pressure at baseline) 
will be compared between arms using the two-stage approach, detailed above.
The first stage will estimate the community-specific proportions of adults known to have HT at baseline who have their HT controlled at follow-up year 3, overall and among those also known to be HIV-positive at baseline (with and without a further restriction on uncontrolled viral replication at baseline). 
For the baseline HIV-HT prevalent population, we will also estimate the community-specific proportions with dual-control at follow-up year 3. 
These proportions will be estimated with TMLE, adjusting for incomplete measures of control at year 3 (including HIV RNA levels for the dual-control outcome); secondary analyses will be unadjusted. 
Primary analyses will be restricted to baseline stable residents; sensitivity analyses will include baseline non-stable residents.
Primary analyses will condition on being alive and resident in the community at follow-up year 3; secondary analyses will further adjust for censoring by death and out-migration.

In the second stage, estimates of control at follow-up year 3 will be compared between arms, weighting individuals equally, and with candidate adjustment variables:
\begin{packed_item}
\item baseline CHC testing coverage and baseline control when estimating the effect on HT control at follow-up year 3 
\item baseline CHC testing coverage and baseline dual-control when estimating the effect on dual-control at follow-up year 3 
\end{packed_item}
We will also test the null hypothesis of no intervention effect stratifying on region (for each population) in unadjusted analyses. We will report estimates of HT control and HT-HIV dual-control stratified by community, by intervention arm, and within intervention by region, sex, and
age (30-44 years, 45-59 years, 60+ years).

\subsubsection{Predictors of uncontrolled HT}\label{HT.predictors}

We will conduct the following analyses to evaluate individual-level predictors of uncontrolled HT at follow-up year 3. 
We will report unadjusted associations and adjusted variable importance measures on the relative scale (statistical analogs of the causal risk ratio), treating each baseline predictor in turn as the intervention variable, and the remainder as the adjustment set. Variable importance measures will be estimated with a pooled individual-level TMLE, adjusted for community as a fixed effect and adjusting for potentially differential missingness of HT control measures. These descriptive statistics and predictor analyses will be reported for the following populations: all adults,  HT-prevalent at follow-up year 3,   HIV-positive at follow-up year 3, and  HIV-HT prevalent at follow-up year 3. For the HIV-HT prevalent population, we will also evaluate predictors of lack of dual-control. 
  Baseline predictors considered will include region, sex, age (30-44 years, 45-49 years, and 60+ years), marital status, education, occupation, household wealth index,  mobility,  alcohol use, and body mass index (BMI). Additional predictors for HIV-positive populations include 
 evidence of prior HIV diagnosis and previous treatment with ART.

\subsubsection{NCD (hypertension and diabetes) control and predictors of uncontrolled NCD}

We will conduct the following analyses to examine the intervention impact on both HT and diabetes (DM) control, as well as predictors of uncontrolled NCD at follow-up year 3. We consider an individual to be NCD prevalent if he or she is HT prevalent and/or DM prevalent at a given time-point. NCD control is defined as current control of all prevalent NCDs. Dual HIV-NCD control is defined as current control of all of prevalent NCDs and plasma HIV RNA level $<$500 copies/ml.

The analyses described in Sections~\ref{HTN3} and \ref{HTN0} will be repeated to estimate NCD control and dual HIV-NCD control at follow-up year 3. 
Specifically, we will estimate the intervention effect on 
\begin{packed_item}
\item NCD control at follow-up year 3 in the overall population of adults with prevalent NCD  at follow-up year 3, in the population of adults who are HIV-NCD prevalent at follow-up year 3, and in the subgroups specified above
\item Dual HIV-NCD control at follow-up year 3 in the population of adults who are HIV-NCD prevalent at follow-up year 3, and within the subgroups specified above
\item NCD control at follow-up year 3 in the overall population of Ugandan adults with prevalent NCD at baseline, in the population of Ugandan adults who are HIV-NCD prevalent at baseline, and within the subgroups specified above 
\item Dual HIV-NCD control at follow-up year 3 in the population of Ugandan adults who are HIV-NCD prevalent at baseline and within the subgroups specified above	
\end{packed_item}
(Recall screening for DM occurred only in Uganda communities at baseline.)
We will also conduct a sensitivity  analysis where all individuals are considered to have blood glucose $\leq$11 mmol/L unless there is evidence otherwise. 

The analyses specified in Section~\ref{HT.predictors} will be repeated to evaluate predictors of uncontrolled NCD at follow-up year 3.

\subsubsection{HT and NCD care cascades}\label{HTNcascade}

Analogously to the HIV care cascade (Section~\ref{cascade.open}),
we will estimate the following population-level metrics at baseline and year 3 ($t=\{0,3\}$): 
prevalence among adult residents (aged $\geq$ 30 years),
the proportion of all prevalent adults who are previously diagnosed,
the proportion of previously diagnosed who have ever initiated treatment, 
the proportion of treatment initiators who are currently controlled,
and the overall population-level control (the proportion of all prevalent adults who are currently controlled). 
These analyses will be performed for the overall population as well as the  HIV-positive population at time $t$. 

Primary analyses will use TMLE to adjust for potentially differential measurement of both disease status and control; unadjusted numbers and proportions will also be reported. 
Primary analyses will include baseline residents (regardless of stability) and in-migrants identified at follow-up year 3; secondary analyses will restrict to baseline stable residents.
Primary analyses will condition on being alive and resident in the community at time $t$; secondary analyses will further adjust for censoring by death and out-migration. 
For both HT (only) and NCD (HT and/or DM), these metrics will be estimated overall, and stratified by community, by intervention arm, and within intervention by region, sex, and
age (30-44 years, 45-59 years, 60+ years). 
 We will generate analogous estimates of dual-control for the HIV-NCD prevalent population. Annual estimates of cascade coverage will also be reported in the intervention arm.

To further understand the impact of the intervention on the HTN and the NCD care cascades, we will conduct the following analyses. 
	\begin{packed_item}
    \item	Using analogous methods to those described in Section~\ref{cascade.open}, we will estimate in each arm  the change in cascade coverage from baseline to follow-up year 3. 
    \item Using analogous methods to those described in Section~\ref{cascade.closed}, we will estimate HT and NCD care cascade outcomes at year 3 in a longitudinal closed cohort of adults with baseline prevalent disease (overall and among baseline HIV-positive). We will characterize this cohort by baseline status (prior diagnosis, treatment and control) as well as evaluate individual-level risk factors for lack of control at follow-up year 3. Within the intervention arm, we will report analogous estimates for years 1 and 2.
	\end{packed_item}

\subsection{Antiretroviral Treatment Associated Toxicities}
We will report number and incidence of grade 3 and grade 4 adverse events and treatment limiting toxicity among individuals initiated on ART outside country guidelines. 

\subsection{Antiretroviral resistance among HIV-positive individuals}
Drug resistance among HIV-infected individuals will be assessed based on the presence of NRTI, PI, and NNRTI mutations. Transmitted resistance will be assessed at year 3 based on the prevalence of resistance mutations at year 3 among  individuals with HIV-seroconversion during the course of the study who remain ART na\"{i}ve at year 3. Transmitted resistance at time of first HIV diagnosis will also be reported among interim seroconversions in the intervention arm of the study. Acquired resistance will be assessed at baseline and year 3 based on the prevalence of resistance mutations among HIV-positive individuals with a history of prior ART initiation, stratified by current ART use at baseline and year 3. 
Resistance will be reported stratified by community, treatment arm, and within treatment arm by region. The demographic characteristics (sex,  age (15-24 years; 25+ years), and  baseline mobility ($<1$mo away; $\ge$1mo away)) of individuals with resistance will be summarized. 
\bibliography{AnalysisPlan.bib}

\begin{thebibliography}{37}
\providecommand{\natexlab}[1]{#1}
\providecommand{\url}[1]{\texttt{#1}}
\expandafter\ifx\csname urlstyle\endcsname\relax
  \providecommand{\doi}[1]{doi: #1}\else
  \providecommand{\doi}{doi: \begingroup \urlstyle{rm}\Url}\fi

\bibitem[Balzer et~al.(2015)Balzer, Petersen, van~der Laan, and the
  {SEARCH}~Consortium]{Balzer2015Adaptive}
L.B. Balzer, M.L. Petersen, M.J. van~der Laan, and the {SEARCH}~Consortium.
\newblock Adaptive pair-matching in randomized trials with unbiased and
  efficient effect estimation.
\newblock \emph{Statistics in Medicine}, 34\penalty0 (6):\penalty0 999--1011,
  2015.
\newblock \doi{10.1002/sim.6380}.

\bibitem[Balzer et~al.(2016{\natexlab{a}})Balzer, Petersen, and van~der
  Laan]{Balzer2016SATE}
L.B. Balzer, M.L. Petersen, and M.J. van~der Laan.
\newblock Targeted estimation and inference of the sample average treatment
  effect in trials with and without pair-matching.
\newblock \emph{Statistics in Medicine}, 35\penalty0 (21):\penalty0 3717--3732,
  2016{\natexlab{a}}.
\newblock \doi{10.1002/sim.6965}.

\bibitem[Hayes and Moulton(2009)]{HayesMoulton2009}
R.J. Hayes and L.H. Moulton.
\newblock \emph{{Cluster Randomised Trials}}.
\newblock Chapman \& Hall/CRC, Boca Raton, 2009.

\bibitem[Balzer et~al.(2016{\natexlab{b}})Balzer, van~der Laan, Petersen, and
  {the SEARCH Collaboration}]{Balzer2016DataAdapt}
L.~Balzer, M.~van~der Laan, M.~Petersen, and {the SEARCH Collaboration}.
\newblock Adaptive pre-specification in randomized trials with and without
  pair-matching.
\newblock \emph{Statistics in Medicine}, 35\penalty0 (10):\penalty0 4528--4545,
  2016{\natexlab{b}}.
\newblock \doi{10.1002/sim.7023}.

\bibitem[Robins(1986)]{Robins1986}
J.M. Robins.
\newblock A new approach to causal inference in mortality studies with
  sustained exposure periods--application to control of the healthy worker
  survivor effect.
\newblock \emph{Mathematical Modelling}, 7:\penalty0 1393--1512, 1986.
\newblock \doi{10.1016/0270-0255(86)90088-6}.

\bibitem[van~der Laan and Rose(2011)]{MarkBook}
M.~van~der Laan and S.~Rose.
\newblock \emph{Targeted Learning: Causal Inference for Observational and
  Experimental Data}.
\newblock Springer, New York Dordrecht Heidelberg London, 2011.

\bibitem[van~der Laan et~al.(2007)van~der Laan, Polley, and
  Hubbard]{SuperLearner}
M.J. van~der Laan, E.C. Polley, and A.E. Hubbard.
\newblock Super learner.
\newblock \emph{Statistical Applications in Genetics and Molecular Biology},
  6\penalty0 (1):\penalty0 25, 2007.
\newblock \doi{10.2202/1544-6115.1309}.

\bibitem[Petersen et~al.(2014)Petersen, Schwab, Gruber, Blaser, Schomaker, and
  van~der Laan]{Petersen2014_JCI}
M.L. Petersen, J.~Schwab, S.~Gruber, N.~Blaser, M.~Schomaker, and M.~van~der
  Laan.
\newblock Targeted maximum likelihood estimation for dynamic and static
  marginal structural working models.
\newblock \emph{Journal of Causal Inference}, 2\penalty0 (2):\penalty0 DOI:
  10.1515/jci--2013--0007, 2014.

\bibitem[van~der Laan et~al.(2012)van~der Laan, Balzer, and
  Petersen]{vanderLaan2012Adaptive}
M.J. van~der Laan, L.B. Balzer, and M.L. Petersen.
\newblock {Adaptive Matching in Randomized Trials and Observational Studies}.
\newblock \emph{Journal of Statistical Research}, 46\penalty0 (2):\penalty0
  113--156, 2012.

\bibitem[Rubin(1990)]{Rubin1990}
Donald~B. Rubin.
\newblock Comment: {N}eyman (1923) and causal inference in experiments and
  observational studies.
\newblock \emph{Statistical Science}, 5\penalty0 (4):\penalty0 472--480, 1990.

\bibitem[Imbens(2004)]{Imbens2004}
G.W. Imbens.
\newblock Nonparametric estimation of average treatment effects under
  exogeneity: a review.
\newblock \emph{Review of Economics and Statistics}, 86\penalty0 (1):\penalty0
  4--29, 2004.
\newblock \doi{10.1162/003465304323023651}.

\bibitem[Imai(2008)]{Imai2008}
K.~Imai.
\newblock Variance identification and efficiency analysis in randomized
  experiments under the matched-pair design.
\newblock \emph{Statistics in Medicine}, 27\penalty0 (24):\penalty0 4857--4873,
  2008.
\newblock \doi{10.1002/sim.3337}.

\bibitem[Fisher(1932)]{Fisher1932}
R.A. Fisher.
\newblock \emph{Statistical methods for research workers}.
\newblock Oliver and Boyd Ltd., Edinburgh, 4th edition, 1932.

\bibitem[Cochran(1957)]{Cochran1957}
W.G. Cochran.
\newblock Analysis of covariance: its nature and uses.
\newblock \emph{Biometrics}, 13:\penalty0 261--281, 1957.
\newblock \doi{10.2307/2527916}.

\bibitem[Cox and Mc{C}ullagh(1982)]{Cox1982}
D.R. Cox and P.~Mc{C}ullagh.
\newblock Some aspects of analysis of covariance.
\newblock \emph{Biometrics}, 38\penalty0 (3):\penalty0 541--561, 1982.
\newblock \doi{10.2307/2530040}.

\bibitem[Tsiatis et~al.(2008)Tsiatis, Davidian, Zhang, and Lu]{Tsiatis2008}
A.A. Tsiatis, M.~Davidian, M.~Zhang, and X.~Lu.
\newblock Covariate adjustment for two-sample treatment comparisons in
  randomized clinical trials: A principled yet flexible approach.
\newblock \emph{Statistics in Medicine}, 27\penalty0 (23):\penalty0 4658--4677,
  2008.
\newblock \doi{10.1002/sim.3113}.

\bibitem[Moore and van~der Laan(2009)]{Moore2009}
K.L. Moore and M.J. van~der Laan.
\newblock Covariate adjustment in randomized trials with binary outcomes:
  {T}argeted maximum likelihood estimation.
\newblock \emph{Statistics in Medicine}, 28\penalty0 (1):\penalty0 39--64,
  2009.
\newblock \doi{10.1002/sim.3445}.

\bibitem[Rosenblum and van~der Laan(2010)]{Rosenblum2010}
M.~Rosenblum and M.J. van~der Laan.
\newblock Simple, efficient estimators of treatment effects in randomized
  trials using generalized linear models to leverage baseline variables.
\newblock \emph{The International Journal of Biostatistics}, 6\penalty0
  (1):\penalty0 Article 13, 2010.
\newblock \doi{10.2202/1557-4679.1138}.

\bibitem[van~der Laan and Rubin(2006)]{vanderLaan2006}
M.J. van~der Laan and D.B. Rubin.
\newblock Targeted maximum likelihood learning.
\newblock \emph{The International Journal of Biostatistics}, 2\penalty0
  (1):\penalty0 Article 11, 2006.
\newblock \doi{10.2202/1557-4679.1043}.

\bibitem[Guwatudde et~al.(2009)Guwatudde, {Wabwire-Mangen}, Eller, Eller,
  McCutchan, Kibuuka, Millard, Sewankambo, Serwadda, Michael, Robb, and {the
  Kayunga Cohort Research Team}]{Guwatudde2009}
D.~Guwatudde, F.~{Wabwire-Mangen}, L.A. Eller, M.~Eller, F.~McCutchan,
  H.~Kibuuka, M.~Millard, N.~Sewankambo, D.~Serwadda, N.~Michael, M.~Robb, and
  {the Kayunga Cohort Research Team}.
\newblock {Relatively Low HIV Infection Rates in Rural Uganda, but with High
  Potential for a Rise: A Cohort Study in Kayunga District, Uganda}.
\newblock \emph{PLoS ONE}, 4\penalty0 (1):\penalty0 e4145, 2009.

\bibitem[Shafer et~al.(2008)Shafer, Biraro, {Nakiyingi-Miiro}, Kamali,
  Ssematimba, Ouma, Ojwiya, Hughes, {Van der Paal}, Whitworth, Opio, and
  Grosskurth]{Shafer2008}
L.A. Shafer, S.~Biraro, J.~{Nakiyingi-Miiro}, A.~Kamali, D.~Ssematimba,
  J.~Ouma, A.~Ojwiya, P.~Hughes, L.~{Van der Paal}, J.~Whitworth, A.~Opio, and
  H.~Grosskurth.
\newblock {HIV} prevalence and incidence are no longer falling in southwest
  {U}ganda: evidence from a rural population cohort 1989-2005.
\newblock \emph{AIDS}, 22\penalty0 (13):\penalty0 1641--1649, 2008.

\bibitem[Gray et~al.(2007)Gray, Kigozi, Serwadda, Makumbi, Watya, Nalugoda,
  Kiwanuka, Moulton, Chaudhary, Chen, Sewankambo, {Wabwire-Mangen}, Bacon,
  Williams, Opendi, Reynolds, Laeyendecker, Quinn, and Wawer]{Gray2007}
R.H. Gray, G.~Kigozi, D.~Serwadda, F.~Makumbi, S.~Watya, F.~Nalugoda,
  N.~Kiwanuka, L.H. Moulton, M.A. Chaudhary, M.Z. Chen, N.K. Sewankambo,
  F.~{Wabwire-Mangen}, M.C. Bacon, C.F. Williams, P.~Opendi, S.J. Reynolds,
  O.~Laeyendecker, T.C. Quinn, and M.J. Wawer.
\newblock Male circumcision for {HIV} prevention in men in {Rakai, Uganda}: a
  randomised trial.
\newblock \emph{Lancet}, 369\penalty0 (9562):\penalty0 657--666, 2007.

\bibitem[Pagel et~al.(2011)Pagel, Prost, Lewycka, Das, Colbourn, Mahapatra,
  Azad, Costello, and Osrin]{Pagel2011intracluster}
C.~Pagel, A.~Prost, S.~Lewycka, S.~Das, T.~Colbourn, R.~Mahapatra, K.~Azad,
  A.~Costello, and D.~Osrin.
\newblock Intracluster correlation coefficients and coefficients of variation
  for perinatal outcomes from five cluster-randomised controlled trials in low
  and middle-income countries: results and methodological implications.
\newblock \emph{Trials}, 12\penalty0 (1):\penalty0 151, 2011.

\bibitem[Hayes et~al.(1995)Hayes, Mosha, Nicoll, Grosskurth, Newell, Killewo,
  Rugemalila, and Mabey]{Hayes1995}
R.~Hayes, F.~Mosha, A.~Nicoll, H.~Grosskurth, J.~Newell, J.~Killewo,
  J.~Rugemalila, and D.~Mabey.
\newblock A community trial of the impact of improved sexually transmitted
  disease treatment on the {HIV} epidemic in rural {Tanzania}: 1. design.
\newblock \emph{{AIDS}}, 9\penalty0 (8):\penalty0 919--926, 1995.

\bibitem[Beck et~al.(2016)Beck, Lu, and Greevy]{nbpMatching}
C.~Beck, B.~Lu, and R.~Greevy.
\newblock \emph{nbp{M}atching: functions for optimal non-bipartite optimal
  matching}, 2016.
\newblock URL \url{https://CRAN.R-project.org/package=nbpMatching}.
\newblock {R} package version 1.5.0.

\bibitem[{Futures Institute}(2014{\natexlab{a}})]{SpectrumManual}
{Futures Institute}.
\newblock \emph{Spectrum manual: {Spectrum} system of policy manuals},
  2014{\natexlab{a}}.

\bibitem[{Futures Institute}(2011)]{GOALSmanual}
{Futures Institute}.
\newblock \emph{Goal Manual - A model for estimating the effects of
  interventions and resource allocation on HIV infections and deaths}, 2011.

\bibitem[{Futures Institute}(2014{\natexlab{b}})]{AIMmanual2014}
{Futures Institute}.
\newblock \emph{{AIM}: A Computer Program for Making {HIV/AIDS} Projections and
  Examining the Demographic and Social Impacts of {AIDS}}, 2014{\natexlab{b}}.

\bibitem[{Uganda Ministry of Health} and {ORC M}acro(2006)]{UHSBS2004}
{Uganda Ministry of Health} and {ORC M}acro.
\newblock {Uganda HIV/AIDS} sero-behavioural survey 2004-2005.
\newblock Calverton, Maryland, USA, 2006.

\bibitem[{Uganda Ministry of Health} and {ICF I}nternational(2012)]{UAIS2011}
{Uganda Ministry of Health} and {ICF I}nternational.
\newblock 2011 {Uganda AIDS} indicator survey: Key findings, Calverton,
  Maryland, USA 2012.

\bibitem[{UNAIDS}(2013)]{UNAIDS2013}
{UNAIDS}.
\newblock {HIV} estimates with uncertainty bounds 1990-2012, 2013.
\newblock Avaliable at
  \url{http://www.unaids.org/en/resources/campaigns/globalreport2013/globalreport/}.

\bibitem[{National AIDS and STI Control Programme (NASCOP)}(2014)]{KAIS2014}
{National AIDS and STI Control Programme (NASCOP)}.
\newblock {Kenya AIDS} indicator survey 2012: Final report.
\newblock Nairobi, {NASCOP}, 2014.

\bibitem[Kimanga and {et al.}(2014)]{Kimanga2014}
D.O. Kimanga and {et al.}
\newblock Prevalence and incidence of {HIV} infection, trends, and risk factors
  among persons aged 15-64 years in {Kenya}: results from a nationally
  representative study.
\newblock \emph{J Acquir Immune Defic Syndr}, 66\penalty0 ({Suppl 1}):\penalty0
  {S13--26}, 2014.

\bibitem[Campbell et~al.(2011)Campbell, Mullins, Hughes, Celum, Wong, Raugi,
  Sorensen, Stoddard, Zhao, Deng, Kahle, Panteleeff, Baeten, McCutchan, Albert,
  Leitner, Wald, Corey, Lingappa, and for the Partners~in Prevention HSV/HIV
  Transmission Study~Team]{Campbell2011}
Mary~S. Campbell, James~I. Mullins, James~P. Hughes, Connie Celum, Kim~G. Wong,
  Dana~N. Raugi, Stefanie Sorensen, Julia~N. Stoddard, Hong Zhao, Wenjie Deng,
  Erin Kahle, Dana Panteleeff, Jared~M. Baeten, Francine~E. McCutchan, Jan
  Albert, Thomas Leitner, Anna Wald, Lawrence Corey, Jairam~R. Lingappa, and
  for the Partners~in Prevention HSV/HIV Transmission Study~Team.
\newblock Viral linkage in hiv-1 seroconverters and their partners in an hiv-1
  prevention clinical trial.
\newblock \emph{PLoS ONE}, 6\penalty0 (3):\penalty0 e16986, 03 2011.
\newblock \doi{10.1371/journal.pone.0016986}.
\newblock URL \url{http://dx.doi.org/10.1371%2Fjournal.pone.0016986}.

\bibitem[Petersen et~al.(2017)Petersen, Balzer, Kwarsiima, Sang,
  et~al.]{PetersenCascade2017}
M.~Petersen, L.~Balzer, D.~Kwarsiima, N.~Sang, et~al.
\newblock Association of implementation of a universal testing and treatment
  intervention with {HIV} diagnosis, receipt of antiretroviral therapy, and
  viral suppression among adults in {East} {Africa}.
\newblock \emph{JAMA}, 317\penalty0 (21):\penalty0 2196--2206, 2017.
\newblock \doi{10.1001/jama.2017.5705}.

\bibitem[Balzer et~al.(2017)Balzer, Schwab, van~der Laan, and
  Petersen]{Balzer2017CascadeMethods}
L.B. Balzer, J.~Schwab, M.J. van~der Laan, and M.L. Petersen.
\newblock Evaluation of progress towards the {UNAIDS} 90-90-90 {HIV} care
  cascade: A description of statistical methods used in an interim analysis of
  the intervention communities in the {SEARCH} study.
\newblock Technical Report 357, University of California at Berkeley, 2017.
\newblock URL \url{http://biostats.bepress.com/ucbbiostat/paper357/}.

\bibitem[Brown et~al.(2016)Brown, Havlir, Ayieko, Mwangwa, Owaraganise,
  et~al.]{Brown2016retention}
L.B. Brown, D.V. Havlir, J.~Ayieko, F.~Mwangwa, A.~Owaraganise, et~al.
\newblock High levels of retention in care with streamlined care and universal
  test and treat in {E}ast {A}frica.
\newblock \emph{{AIDS}}, 30:\penalty0 2855--2964, 2016.

\end{thebibliography}

\end{document}